\documentclass[twocolumn,showpacs,preprintnumbers,prb,aps,psfig]{revtex4}
\usepackage{graphicx}
\newcommand{\be}{\begin{equation}}
\newcommand{\ee}{\end{equation}}
\newcommand{\bea}{\begin{eqnarray}}
\newcommand{\eea}{\end{eqnarray}}

\begin{document}
\title{Large Miscibility Gap in the Ba(Mn$_x$Fe$_{1-x}$)$_2$As$_2$ System}
\author {Abhishek Pandey, V. K. Anand, and D. C. Johnston} 
\affiliation {Ames Laboratory and Department of Physics and Astronomy, Iowa State University, Ames, Iowa 50011}


\date{\today}

\begin{abstract}

The compounds BaMn$_2$As$_2$ and BaFe$_2$As$_2$ both crystallize in the body-centered-tetragonal ThCr$_2$Si$_2$-type (122-type) structure at room temperature but exhibit quite different unit cell volumes and very different magnetic and electronic transport properties.  Evidently reflecting these disparities, we have discovered a large miscibility gap in the system Ba(Mn$_x$Fe$_{1-x}$)$_2$As$_2$.  Rietveld refinements of powder x-ray diffraction (XRD) measurements on samples slow-cooled from 1000\,$^\circ$C to room temperature (RT) reveal a two-phase mixture of BaMn$_2$As$_2$ and ${\rm Ba(Mn_{0.12}Fe_{0.88})_2As_2}$ phases together with impurity phases for $x = 0.2$, 0.4, 0.5, 0.6 and 0.8.  We infer that there exists a miscibility gap in this system at 300~K with composition limits $0.12 \lesssim x \lesssim 1$.  For samples quenched from 1000\,$^\circ$C to 77~K, the refinements of RT XRD data indicate that the miscibility gap at RT narrows at 1000\,$^\circ$C to $0.2 \lesssim x\lesssim 0.8$.  Samples with $x=0.4$, 0.5 and 0.6 quenched from 1100--1400~$^\circ$C to 77~K contain a single 122-type phase together with significant amounts of Fe$_{1-x}$Mn$_x$As and FeAs$_2$ impurity phases. These results indicate that the system is not a pseudo-binary system over the whole composition range and that the 122-type phase has a significant homogeneity range at these temperatures.  Magnetic susceptibility $\chi$, electrical resistivity $\rho$ and heat capacity measurements versus temperature $T$ of the single-phase quenched polycrystalline samples with $x=0.2$ and 0.8 are reported.  We also report attempts to grow single crystals of the substituted compounds Ba(Mn$_{1-x}T_x$)$_2$As$_2$ ($T$ = Cr, Fe, Co, Ni, Cu, Ru, Rh, Pd, Re, Pt) and BaMn$_2$(As$_{1-x}$Sb$_x$)$_2$ out of Sn flux. Energy dispersive x-ray analyses show that most of these elements do not substitute into the lattice in amounts greater than 0.5\%. However, concentrations of 4.4\%, $\sim10$\% and 2.6\% were achieved for Cr, Fe and Sb substitutions, respectively, and $\chi(T)$ and $\rho(T)$ data for these crystals are presented.

\end{abstract}


\maketitle

\section{INTRODUCTION}

The discovery\cite{Kamihara1} of superconductivity at 26~K in LaFeAsO$_{1-x}$F$_x$ with the tetragonal ZrCuSiAs-type structure in 2008 motivated many efforts to understand the superconducting pairing mechanism and to discover additional families of Fe-based superconductors. \cite{Johnston2010} New parent compounds $A$Fe$_2$As$_2$ ($A$ = Ba, Sr and Ca) that crystallize in the related tetragonal ThCr$_2$Si$_2$-type (122-type) structure, together with other families, were soon discovered.\cite{Johnston2010}  Among 122-type materials, BaFe$_2$As$_2$ has gained much attention as a parent compound because superconductivity can be induced in this material by various means, \emph{i.e.}, by applying pressure \cite{Yamazaki1, Ishikawa1, Colombier1} and by substitutions at the Ba site (by K)\cite{Rotter1}, at the Fe site (by Co, Ni, Ru, Rh, and Pd),\cite{Johnston2010, Canfield2010} and at the As site (by P).\cite{Jiang1} An interesting result is that while electron doping by Co, Ni, Rh and Pd at the Fe site in BaFe$_2$As$_2$ induces superconductivity, hole doping by Cr and Mn for Fe does not.\cite{Sefat1, Liu2010, Budko1, Kim2010b}

The reported experimental results and electronic structure calculations demonstrate that the physical properties of BaCr$_2$As$_2$ and BaMn$_2$As$_2$ are significantly different from those of BaFe$_2$As$_2$. \cite{DJSingh1, An1, YSingh1, YSingh2} In particular, the unit cell volume of BaMn$_2$As$_2$  (234.12~\AA$^3$) is much larger than that of BaFe$_2$As$_2$ (204.38~\AA$^3$),\cite{Johnston2010} in spite of the fact that Mn and Fe are 3$d$ transition-metal neighbors in the Periodic Table. This divergence suggests that the Mn$^{+2}$ is in a high-spin state while, in a local moment model, Fe$^{+2}$ is in a low-spin state.  ${\rm BaMn_2As_2}$ is an antiferromagnetic insulator with a large ordered moment of $3.9\,\mu_{\rm B}$/Mn and a high N\'eel temperature $T_{\rm N} = 625$~K (Ref.~\onlinecite{YSingh2}) whereas BaFe$_2$As$_2$ is an itinerant antiferromagnet with a much lower ordered moment of $0.9\,\mu_{\rm B}$/Fe and a much lower $T_{\rm N} = 137$~K\@.\cite{Johnston2010}  In addition, the antiferromagnetic structure of BaMn$_2$As$_2$ is N\'eel-type (G-type or checkerboard-type) with the ordered moments aligned along the tetragonal $c$~axis,\cite{YSingh2} whereas that of BaFe$_2$As$_2$ is a type of antiferromagnetic stripe structure with the ordered moments aligned in the $ab$~plane.\cite{Johnston2010}

There have been a number of investigations of iron-arsenide compounds in which the Fe is partially substituted by Mn.\cite{Kim2010b, Cheng2010, Kim2010a, Kim2010c,  Kasinathan2009, Liu2010, Berardan2009, Huang2010}  For $A\rm{Fe_2As_2}$-type materials, a maximum of $17.6\%$ and $18\%$ Fe substitution by Mn was reported for  $\rm{BaFe_2As_2}$ (Ref.~\onlinecite{Kim2010b}) and $\rm{SrFe_2As_2}$ (Ref.~\onlinecite{Kim2010a}), respectively. Although the Mn doping leads to suppression of the magnetic and structural transitions to lower temperatures, the reported results suggest that doping with Mn is detrimental to the occurrence of superconductivity and in most cases it leads to a magnetic and more resistive ground state.\cite{Cheng2010, Kim2010a, Kim2010b, Kim2010c, Kasinathan2009, Liu2010, Berardan2009} It has been reported that in hole-doped systems, besides the changes in the electronic structure introduced by the dopant, the resultant variation of the Fe-As bond length plays a crucial role in defining the nature of the ground state.\cite{Kim2010a, Kasinathan2009} The variation of Fe-As bond length for K- and Mn-doped $\rm{SrFe_2As_2}$ is significantly different.\cite{Kim2010a}  In none of these studies of $A$(Mn$_x$Fe$_{1-x}$)$_2$As$_2$ systems was the possible existence of a miscibility gap in the phase diagram mentioned.

The peculiar properties of ${\rm BaMn_2As_2}$ suggest that this compound may be a bridge between the BaFe$_2$As$_2$-type materials and the more ionic high-$T_{\rm c}$ cuprates.\cite{YSingh1, YSingh2}  This in turn suggests that if ${\rm BaMn_2As_2}$ could be doped into the metallic state, the resulting compounds might have interesting superconducting properties.  This possibility motivated our attempts to grow single crystals of substituted ${\rm BaMn_2As_2}$, as will be discussed.  However, we found that the attempted substitutions usually failed to achieve concentrations above 0.5\%.  In the case of Fe substitutions for Mn, the single crystals contained a maximum of $\sim 10$\% Fe substituted for Mn.  It was not clear to us whether such limits arose from peculiarities of the crystal growths, or were intrinsic to the equilibrium phase diagrams.  To investigate this question, we chose to study Ba(Mn$_x$Fe$_{1-x}$)$_2$As$_2$ in more detail as a prototype system using polycrystalline samples.  We discovered a wide miscibility gap in this system in slow-cooled samples, which only narrows slightly for samples quenched from 1000~$^\circ$C to 77~K\@.  Magnetic susceptibility $\chi$, electrical resistivity $\rho$ and heat capacity $C_{\rm p}$ versus temperature $T$ data for single-phase polycrystalline ${\rm Ba(Mn_{0.2}Fe_{0.8})_2As_2}$ and ${\rm Ba(Mn_{0.8}Fe_{0.2})_2As_2}$ samples quenched from 1000~$^\circ$C will be reported.  We also present $\rho(T)$ and/or $\chi(T)$ data for several of the above-mentioned lightly-doped crystals.

The remainder of the paper is organized as follows.  The experimental details are given in Sec.~\ref{ExpDetails}. The phase analyses and structures of polycrystalline and single crystal samples with nominal composition Ba(Mn$_x$Fe$_{1-x}$)$_2$As$_2$, which establish the miscibility gap, are given in Sec.~\ref{SecMiscGap}.  The physical properties of the single-phase polycrystalline ${\rm Ba(Mn_{0.2}Fe_{0.8})_2As_2}$ and ${\rm Ba(Mn_{0.8}Fe_{0.2})_2As_2}$ samples quenched from 1000~$^\circ$C are presented in Sec~\ref{SecPowderProps}.  Our $\chi(T)$ and $\rho(T)$ data for  substituted single crystals of ${\rm BaMn_2As_2}$ are presented in Sec.~\ref{SecXtalResults}.  A summary of this work and our conclusions are given in Sec.~\ref{SummConcl}.

\section{\label{ExpDetails} EXPERIMENTAL DETAILS}

Polycrystalline samples of Ba(Mn$_x$Fe$_{1-x}$)$_2$As$_2$ ($x=0.2$, 0.4, 0.5, 0.6 and 0.8) were synthesized by solid state reaction using Ba (99.99\%) from Sigma-Aldrich, and Fe (99.998\%), Mn (99.95\%) and As (99.99999\%) from Alfa-Aesar.  Stoichiometric mixtures of the elements were pelletized and placed in alumina crucibles that were sealed inside evacuated silica tubes. The samples were heated to 585\,$^\circ$C at a rate of 40\,$^\circ$C/h, held there for 20\,h, then heated to 620\,$^\circ$C at a rate of 7\,$^\circ$C/h, held there for 20\,h, and finally heated to 750\,$^\circ$C at 30\,$^\circ$C/h and held there for 25\,h. The samples were then thoroughly ground, pelletized and again sealed in evacuated silica tubes. The samples were fired at 750\,$^\circ$C and then at 1000\,$^\circ$C for 20~h and 80~h, respectively, followed by furnace-cooling (``slow-cooling'').  For quenching experiments, pressed pellets were wrapped in Ta foil and sealed in silica tubes containing $\sim 1/3$\,atm of Ar. The tubes were then heated to temperatures ranging from  1000 to 1400\,$^\circ$C for 2--20~h and then quenched into liquid nitrogen at 77~K\@.  There was no evidence that the samples reacted with either the Ta foils or the silica tubes during these quenching experiments.

Structural characterization of the polycrystalline samples was performed using powder x-ray diffraction (XRD) data obtained from a Rigaku Geigerflex powder diffractometer and CuK$_\alpha$ radiation. The FullProf package was used for Rietveld refinement of the XRD data.\cite{Carvajal1}

Attempts to grow single crystals of Ba(Mn$_{1-x}T_x$)$_2$As$_2$ ($T$ = Cr, Fe, Co, Ni, Cu, Ru, Rh, Pd, Re and Pt) and BaMn$_2$(As$_{1-x}$Sb$_x$)$_2$ were made using Sn flux. To begin, polycrystalline samples of Ba(Mn$_{1-x}T_x$)$_2$As$_2$ or BaMn$_2$(As$_{1-x}$Sb$_x$)$_2$ were synthesized using the method described above and then $\sim 200$~mg of a sample was placed in an alumina crucible together with Sn, using a sample to Sn molar ratio of $1:45$. The top of the crucible was packed with quartz wool. The alumina crucible was sealed inside a silica tube under a pressure of $\sim$1/3~atm~Ar. The sample was heated to 1100\,$^\circ$C at a rate of 40\,$^\circ$C/h, held there for 40 h, and then cooled to either 700\,$^\circ$C or 900\,$^\circ$C at a rate of 4\,$^\circ$C/h. At the respective temperature the Sn flux was decanted using a centrifuge. The typical size of the shiny plate-like crystals obtained was $2 \times 2 \times 0.2 $ mm$^3$.

\begin{table}
\caption{\label{Table:EDX} Measured values of the content of transition metals ($T$ = Cr, Fe, Co, Ni, Cu, Ru, Rh, Pd, Re and Pt) and Sb present in the Sn-grown Mn- and As-rich crystals of Ba(Mn$_{1-x}$$T_{x}$)$_2$As$_2$ and BaMn$_2$(As$_{1-x}$Sb$_{x}$)$_2$, respectively.  A value of ``0'' means that the EDX measurement software gave this value with no error bars.  The Sn content is the atomic fraction with respect to the total transition metal content. The Fe-substituted sample with high Fe concentration marked by an asterisk is Crystal-1 in Table~\ref{Table:50:50 Mn-Fe Xtals} below, for which the Fe and Sn concentrations at two different points on the crystal surface were measured.}

 \begin{ruledtabular}
		\begin{tabular}{l c c c}
		 Substitutions & Site & Amount(\%) & Sn-content (\%) \\
	 \hline
			
			 Cr & Mn-site & $4.4\pm0.9$ & $0.7\pm0.5$ \\
			 Fe & Mn-site & $9.9\pm0.5$ & $0.6\pm0.1$ \\
			 Fe* & Mn-site & $91.3\pm0.9$ & $2.7\pm0.3$ \\
				& 		& $91.6\pm0.9$ & $1.9\pm0.1$ \\
			 Co & Mn-site & $0.2\pm0.3$ & $0.2\pm0.1$ \\
			 Ni & Mn-site & $0.1\pm0.2$ & ``0''         \\
			 Cu & Mn-site & $0.5\pm0.3$ & $0.5\pm0.1$ \\
			 Ru & Mn-site & $0.2\pm0.9$ & ``0''         \\
			 Rh & Mn-site & $0.2\pm0.3$ & ``0''         \\
			 Pd & Mn-site & ``0"         & $0.4\pm0.1$ \\
			 Re & Mn-site & ``0"         & $0.3\pm0.3$ \\
			 Pt & Mn-site & ``0"         & $0.5\pm0.3$ \\
			 Sb & As-site & $2.6\pm0.3$ & $0.3\pm0.2$ \\
	            
		\end{tabular}
		\end{ruledtabular}
\end{table}

The chemical compositions of the crystals grown in Sn flux were measured by energy dispersive x-ray (EDX) analysis using a JEOL scanning electron microscope. The EDX results showed that most growth attempts resulted in substitutions of less than 0.5\% except for the cases of Cr, Fe and Sb where 4.4\%, $\sim 10\%$ and 2.6\% substitutions were obtained, respectively, as shown in Table~\ref{Table:EDX}.  A more detailed study of the compositions of Fe-substituted crystals is given in Sec.~\ref{Sec:MnFeXtal_Compositions}.

An interesting result is that in contrast to the case of BaFe$_2$As$_2$ crystals grown in Sn flux where up to 2--3~at\% of Sn  with respect to Fe becomes incorporated into the crystals\cite{Ni2008} that drastically modifies their physical properties,\cite{Baek2008, Wang2009, Su2009} Sn enters the crystal lattice of pure or lightly substituted BaMn$_2$As$_2$ in much smaller amounts (see Table~\ref{Table:EDX}).  Consistent with this result, our lightly Cr-, Fe- and Sb-substituted crystals were found to have magnetic properties very similar to those of pure BaMn$_2$As$_2$, as will be shown in Sec.~\ref{SecXtalResults}.

Field- and temperature-dependent magnetization measurements were carried out using a superconducting quantum interference device magnetometer (Quantum Design, Inc.).  Heat capacity measurements were done using a thermal relaxation technique with a Quantum Design physical property measurement system (PPMS).

Electrical resistivity measurements were carried out using a four-probe ac method in the PPMS\@.  Contacts of the 0.05~mm diameter Pt electrical leads to the samples were made using silver epoxy for the polycrystalline samples and ${\rm In_{98}Ag_2}$ solder for the crystals.  The measurement current amplitude and frequency were 1~mA and 38~Hz for the polycrystalline samples, and 0.01~mA and 38~Hz and 18~Hz for the Fe- and Cr-doped crystals, respectively.  The room-temperature contact resistances were 4--5~$\Omega$ for the quenched polycrystalline ${\rm Ba(Mn_{0.8}Fe_{0.2})_2As_2}$  and ${\rm Ba(Mn_{0.2}Fe_{0.8})_2As_2}$ samples and 0.1 and 1~k$\Omega$ for the Cr- and Fe-doped ${\rm BaMn_2As_2}$ crystals, respectively.  The validity of the resistivity data was verified by examining the voltage waveform at representative temperatures for each sample.

\section{\label{SecMiscGap} Miscibility Gap in the B\lowercase{a}(M\lowercase{n}$_x$F\lowercase{e}$_{1-x})_2$A\lowercase{s}$_2$ System}

\begin{figure}
	\includegraphics[width=3in]{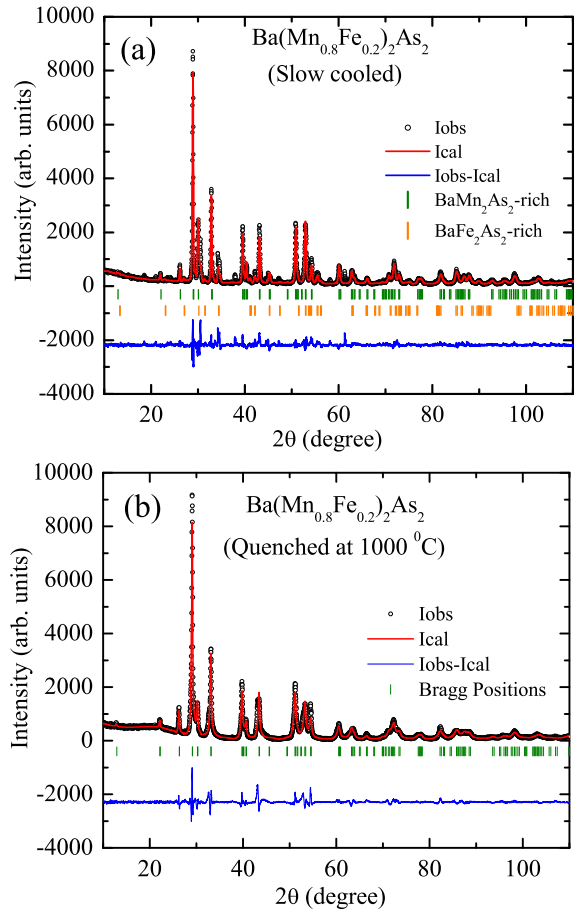}
	\caption{(Color online)  (a) Room-temperature powder XRD pattern (open circles) of a slow-cooled sample of Ba(Mn$_{0.8}$Fe$_{0.2}$)$_2$As$_2$. The solid line is a two-phase (BaMn$_2$As$_2$-rich and BaFe$_2$As$_2$-rich phases) Rietveld refinement fit. Difference profile and Bragg positions of both phases are also shown. (b) Room temperature powder XRD pattern (open circles) of a quenched (1000~$^\circ$C to 77~K) sample of Ba(Mn$_{0.8}$Fe$_{0.2}$)$_2$As$_2$, along with the Rietveld fit, difference profile and Bragg peak positions.}
	\label{fig:Figure_1_AP1066_XRD}
\end{figure}

\begin{figure}
	\includegraphics[width=3in]{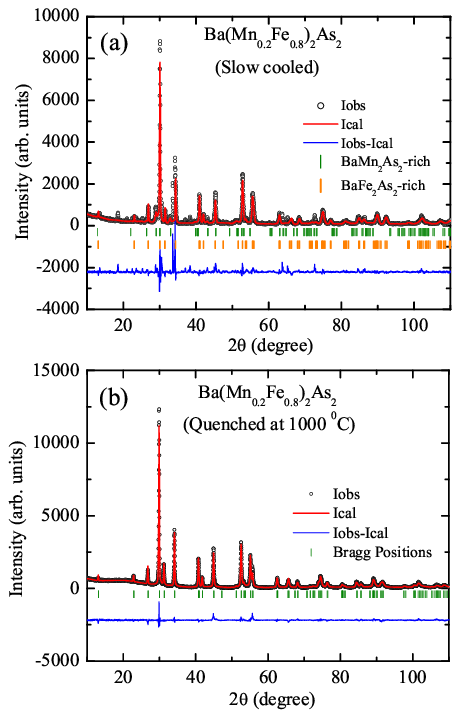}
	\caption{(Color online)  (a) Room temperature powder XRD pattern (open circles) of a slow-cooled sample of Ba(Mn$_{0.2}$Fe$_{0.8}$)$_2$As$_2$.  The solid line is a two-phase (BaMn$_2$As$_2$-rich and BaFe$_2$As$_2$-rich phases) Rietveld refinement.  The difference profile and Bragg peak positions of both phases are also shown.  (b) Room-temperature powder XRD pattern (open circles) of a single-phase sample of Ba(Mn$_{0.2}$Fe$_{0.8}$)$_2$As$_2$ obtained by quenching from 1000\,$^\circ$C to 77~K, along with the Rietveld refinement fit, difference profile and Bragg peak positions.}
	\label{fig:Figure_2_AP1069_XRD}
\end{figure}

\begin{table*}
\caption{\label{Table:CrystalData}  Crystal data at room temperature for the 122-type phase(s) in slow-cooled (SC) and quenched (Q) phases of polycrystalline Ba(Mn$_x$Fe$_{1-x}$)$_2$As$_2$ ($x$ = 0.2, 0.4, 0.5, 0.6 and 0.8).  All samples except for the quenched ${\rm Ba(Mn_{0.2}Fe_{0.8})_{2}As_{2}}$ and ${\rm Ba(Mn_{0.8}Fe_{0.2})_{2}As_{2}}$ samples also contained impurity phases.  The temperature in parentheses following a ``Q'' symbol is the temperature from which the sample was quenched.  The crystal data were  obtained from Rietveld refinements of powder XRD data in the body-centered-tetragonal space group $I$4/$mmm$ using the atomic positions of the ThCr$_2$Si$_2$-type structure.\cite{Johnston2010} The tetragonal lattice parameters are $a$ and $c$ and $V_{\rm cell}$ is the unit cell volume.  $M$ represents transition metal atoms (Mn and Fe), $z_{\rm As}$ is the $c$-axis position parameter of As, $d_{M-{\rm As}}$ is the Mn/Fe-As bond distance and $\theta_{2}$ is the twofold As-$M$-As bond angle\cite{Johnston2010} within a distorted $M$-centered $M$As$_4$ tetrahedron.}
\begin{ruledtabular}
		\begin{tabular}{l c c c c c c c }
		 Compound & SC or Q & $a$   & $c$   & $V_{\rm cell}$       & $z_{\rm As}$ & $d_{M-{\rm As}}$ & $\theta_{2}$ \\
		           &      & (\AA) & (\AA) & (\AA$^3$) &       & (\AA)     & (deg)        \\
		  
		  \hline
			${\rm Ba(Mn_{0.2}Fe_{0.8})_{2}As_{2}}$ & Q (1000 $^\circ{\rm C}$) & 4.0014(1) & 13.1195(5) & 210.06(1) & 0.3566(1) & 2.441(2) & 110.1(1) \\
			                                       & SC (Phase-1)             & 4.169(1)  & 13.432(8)  & 233.4(2)  & 0.358(1)  & 2.54(2)  & 110(1)   \\
			                                       & SC (Phase-2)             & 3.9676(3) & 13.071(1)  & 205.76(3) & 0.3550(2) & 2.412(3) & 110.7(2) \\
		  ${\rm Ba(Mn_{0.4}Fe_{0.6})_{2}As_{2}}$ & Q (1100 $^\circ{\rm C}$) & 4.0565(4) & 13.245(2)  & 217.95(5) & 0.3577(2) & 2.480(4) & 109.8(2) \\
		                                         & SC (Phase-1)             & 4.1716(3) & 13.481(1)  & 234.60(3) & 0.3619(3) & 2.574(5) & 108.3(3) \\
		                                         & SC (Phase-2)             & 3.9837(2) & 13.0770(7) & 207.53(2) & 0.3562(2) & 2.428(2) & 110.2(2) \\
		  ${\rm Ba(Mn_{0.5}Fe_{0.5})_{2}As_{2}}$ & Q (1400 $^\circ{\rm C}$) & 4.1135(5) & 13.340(2)  & 225.74(5) & 0.3599(2) & 2.526(4) & 109.0(2) \\
		                                         & SC (Phase-1)             & 4.1659(2) & 13.4679(7) & 233.74(2) & 0.3614(2) & 2.567(3) & 108.5(2) \\
		                                         & SC (Phase-2)             & 3.9744(2) & 13.0605(7) & 206.30(2) & 0.3554(1) & 2.417(2) & 110.6(1) \\
		  ${\rm Ba(Mn_{0.6}Fe_{0.4})_{2}As_{2}}$ & Q (1250 $^\circ{\rm C}$) & 4.099(1)  & 13.313(4)  & 223.7(1)  & 0.3584(3) & 2.507(6) & 109.7(3) \\
		                                         & SC (Phase-1)             & 4.1719(3) & 13.484(1)  & 234.68(3) & 0.3620(2) & 2.575(4) & 108.2(2) \\
		                                         & SC (Phase-2)             & 3.9863(3) & 13.073(1)  & 207.74(3) & 0.3555(3) & 2.424(5) & 110.6(3) \\
		  ${\rm Ba(Mn_{0.8}Fe_{0.2})_{2}As_{2}}$ & Q (1000 $^\circ{\rm C}$) & 4.1368(3) & 13.388(1)  & 229.11(3) & 0.3615(1) & 2.551(2) & 108.4(1) \\
		                                         & SC (Phase-1)             & 4.1697(3) & 13.4714(9) & 234.22(3) & 0.3626(2) & 2.578(4) & 107.9(2) \\
		                                         & SC (Phase-2)             & 3.9880(5) & 13.043(3)  & 207.43(6) & 0.3567(5) & 2.432(8) & 110.2(5) \\
		\end{tabular}
\end{ruledtabular}
\end{table*}

\begin{figure}
	\includegraphics[width=2.69in]{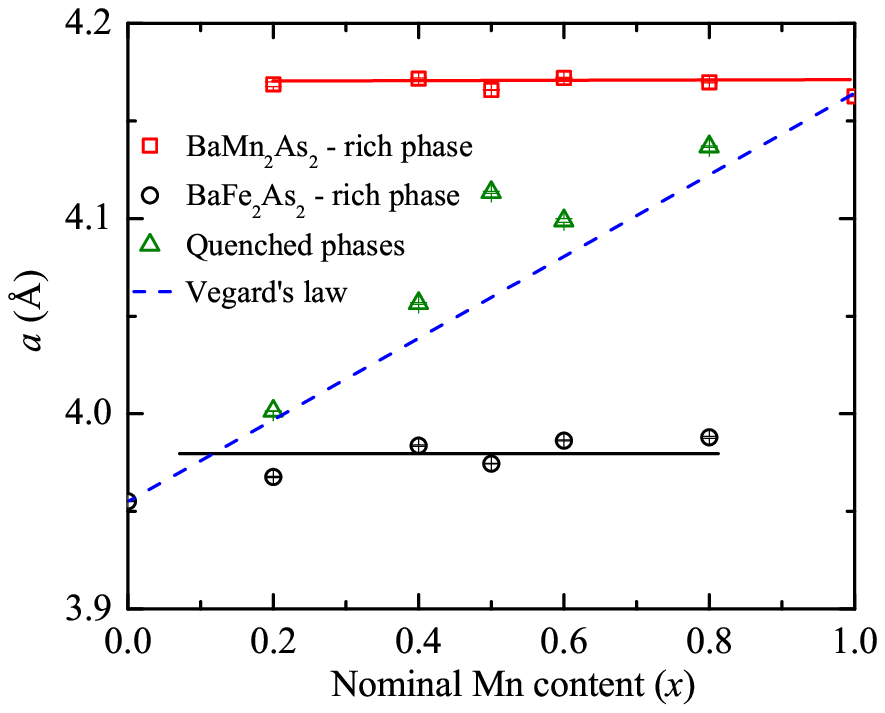}
 	\includegraphics[width=2.69in]{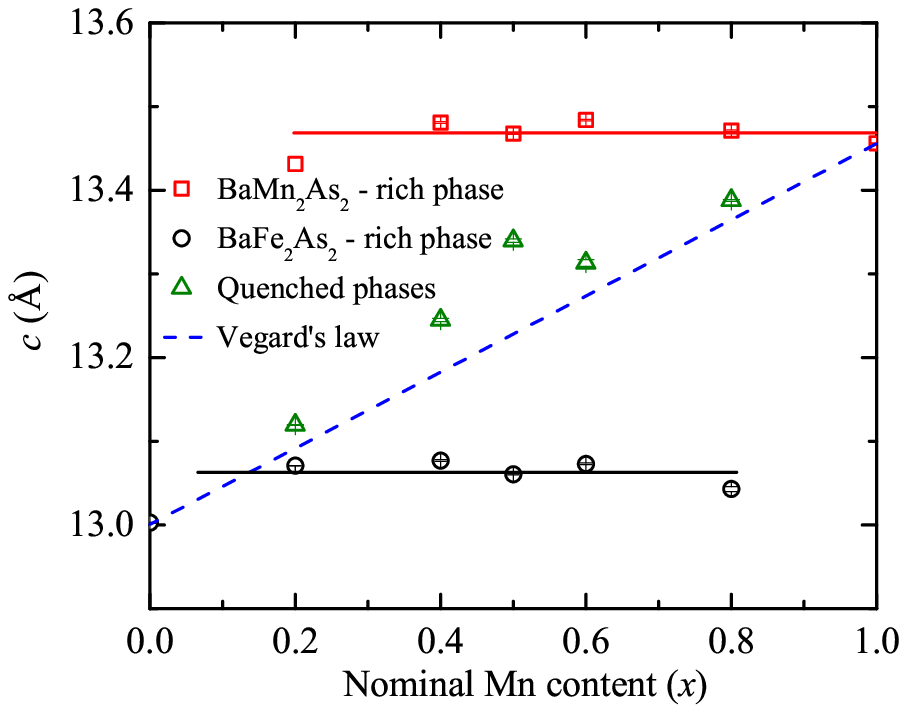}
	\includegraphics[width=2.69in]{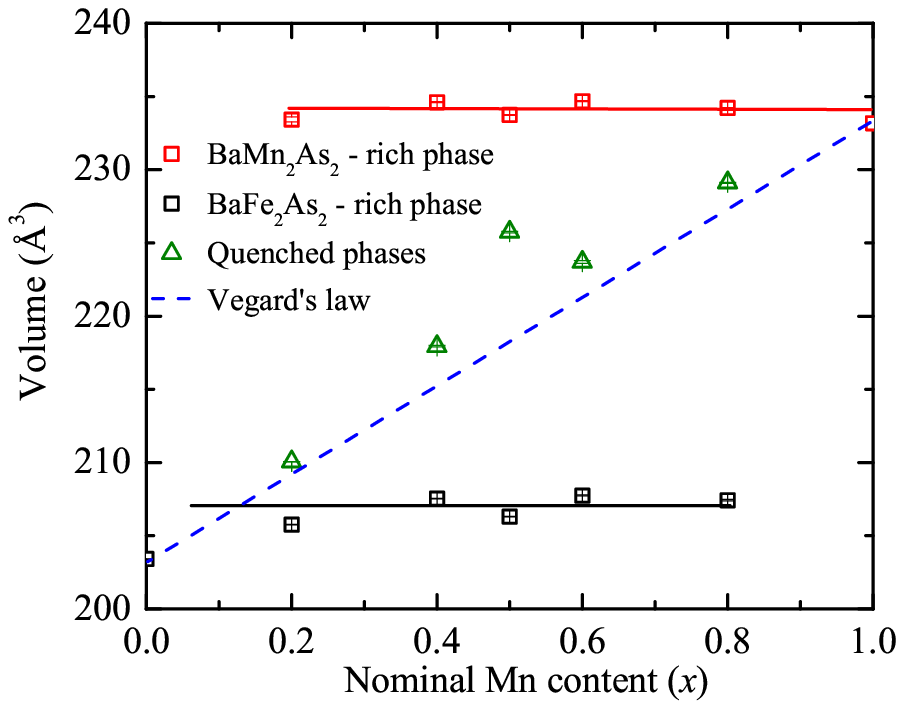}
	\caption{(Color online) Variations of the tetragonal lattice parameters $a$ and $c$ and of the unit cell volumes for the BaMn$_2$As$_2$-rich and BaFe$_2$As$_2$-rich phases present in slow-cooled polycrystalline samples of Ba(Mn$_x$Fe$_{1-x}$)$_2$As$_2$ versus the nominal Mn concentration $x$.  These data indicate the presence of a misibility gap in the composition range $0.12\lesssim x \lesssim 1$ at 300~K\@.  Also shown are data for quenched samples, within which only the $x=0.2$ and $x=0.8$ samples were single-phase.  Error bars are appended to the data points and are smaller than the size of the symbols.}
	\label{fig:Figure_Lattice_Parameters}
\end{figure}

\begin{figure}
	\includegraphics[width=3in]{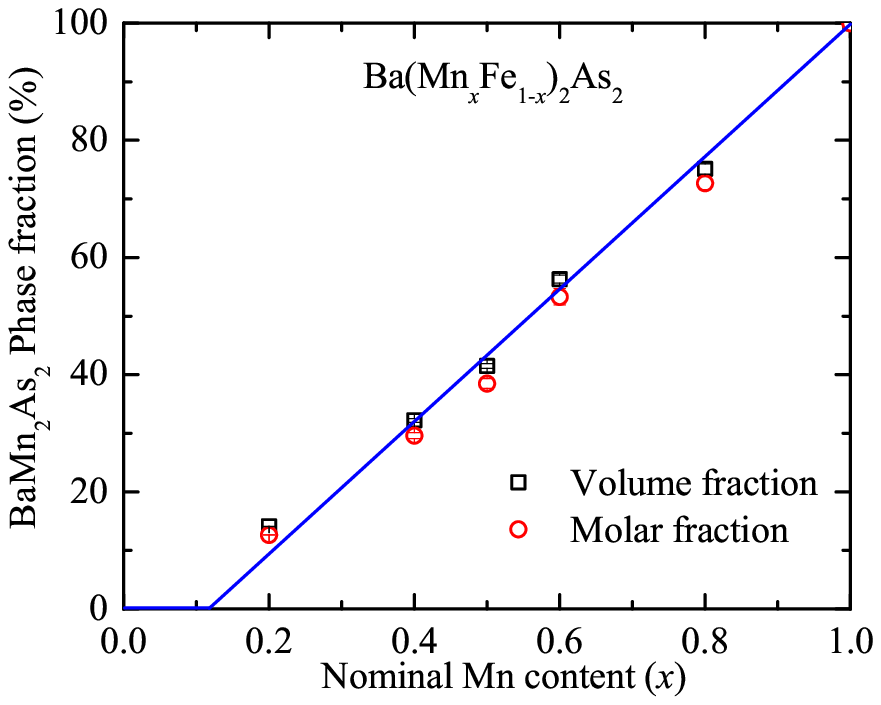}
	\caption{(Color online) Volume and molar fractions of the BaMn$_2$As$_2$-rich phase as a function of nominal Mn concentration $x$ in slow-cooled samples of polycrystalline Ba(Mn$_x$Fe$_{1-x}$)$_2$As$_2$.  The straight-line fit indicates that the miscibility gap at 300~K has the composition range $0.12 \lesssim x \lesssim 1$, consistent with the crystal data in Fig.~\ref{fig:Figure_Lattice_Parameters}.}
	\label{fig:Figure_Molar_Fraction}
\end{figure}

\subsection{\label{SecSlowCooled} Slow-Cooled Polycrystalline Samples}

A room-temperature XRD pattern of a slow-cooled sample of Ba(Mn$_{0.8}$Fe$_{0.2}$)$_2$As$_2$ is  shown in Fig.~\ref{fig:Figure_1_AP1066_XRD}(a).  The XRD data for the sample indicate that the sample contains a significant amount of impurities, mainly Fe$_{1-x}$Mn$_x$As. We refined the XRD data using a two-phase (BaMn$_2$As$_2$-rich and BaFe$_2$As$_2$-rich phases) Rietveld refinement but did not include the impurity phases in the refinement. For each 122-type phase, the data were refined in the body-centered-tetragonal space group $I$4/$mmm$ using the atomic positions of the ThCr$_2$Si$_2$-type structure.\cite{Johnston2010}   Figure~\ref{fig:Figure_1_AP1066_XRD}(a) shows the Rietveld fit, the difference profile between the data and fit, and the fitted line positions.  The room temperature powder XRD pattern of the slow-cooled sample of Ba(Mn$_{0.2}$Fe$_{0.8}$)$_2$As$_2$ is shown in Fig.~\ref{fig:Figure_2_AP1069_XRD}(a). Similar to the case of Ba(Mn$_{0.8}$Fe$_{0.2}$)$_2$As$_2$ in Fig.~\ref{fig:Figure_1_AP1066_XRD}(a), the slow-cooled sample of Ba(Mn$_{0.2}$Fe$_{0.8}$)$_2$As$_2$ contains a significant amount of impurities. A two-phase Rietveld refinement of the data was carried out in the same way as for Ba(Mn$_{0.8}$Fe$_{0.2}$)$_2$As$_2$ to obtain the unit cell parameters and the volumes and percentage molar fractions of BaMn$_2$As$_2$-rich and BaFe$_2$As$_2$-rich phases.  This fit is shown in Fig.~\ref{fig:Figure_2_AP1069_XRD}(a).

Similar to the above cases of Ba(Mn$_{0.8}$Fe$_{0.2}$)$_2$As$_2$ and Ba(Mn$_{0.2}$Fe$_{0.8}$)$_2$As$_2$, the slow-cooled samples of Ba(Mn$_{1-x}$Fe$_x$)$_2$As$_2$ (\emph{x} =  0.4, 0.5 and 0.6) also contain phase-separated mixtures of the Fe-rich and Mn-rich 122-type phases together with impurity phases.   We carried out similar two-phase Rietveld refinements of the XRD patterns of these samples for the Mn-rich and Fe-rich 122-type phases.  The $a$ and $c$ lattice parameters and the unit cell volumes of the BaMn$_2$As$_2$-rich and BaFe$_2$As$_2$-rich phases in Ba(Mn$_x$Fe$_{1-x}$)$_2$As$_2$ are listed in Table~\ref{Table:CrystalData} and plotted versus nominal Mn content $x$ for all five samples in Fig.~\ref{fig:Figure_Lattice_Parameters}, together with data\cite{Johnston2010} for $x = 0$ and $x = 1$ from the literature.    The corresponding volume and molar fraction of the BaMn$_2$As$_2$-rich phase in the samples are plotted versus nominal Mn concentration $x$ in Fig.~\ref{fig:Figure_Molar_Fraction}.  Since impurity phases were not accounted for in the refinements, the values of the molar fraction in Fig.~\ref{fig:Figure_Molar_Fraction} represent the relative fractions of the BaMn$_2$As$_2$-rich and BaFe$_2$As$_2$-rich phases.

Considering Vegard's law shown in Fig.~\ref{fig:Figure_Lattice_Parameters} and the straight-line fit to the data in Fig.~\ref{fig:Figure_Molar_Fraction}, the data in these two figures consistently indicate that a miscibility gap occurs in the Ba(Mn$_x$Fe$_{1-x}$)$_2$As$_2$ system at 300~K (for the slow-cooled samples) that has the composition range $0.12 \lesssim x \lesssim 1$.  That is, about 12\% of Mn can be substituted for Fe in BaFe$_2$As$_2$ but almost no Fe can be substituted for Mn in BaMn$_2$As$_2$.  In the following section we investigate whether the miscibility gap boundaries are temperature-dependent.

\subsection{Quenched Polycrystalline Samples}

The Rietveld refinement of the powder XRD data for a Ba(Mn$_{0.8}$Fe$_{0.2}$)$_2$As$_2$ sample quenched  from 1000\,$^\circ$C to 77~K, shown in Fig.~\ref{fig:Figure_1_AP1066_XRD}(b), indicates that the sample is a single phase, with no evidence of phase separation of different compositions of 122-type phases or of impurity phases. The unit cell volume calculated from the lattice parameters is 229.11(3)~\AA$^3$, which is 2.19\% smaller than that of pure BaMn$_2$As$_2$. Similarly, Rietveld refinement of a quenched Ba(Mn$_{0.2}$Fe$_{0.8}$)$_2$As$_2$ sample shown in Fig.~\ref{fig:Figure_2_AP1069_XRD}(b) exhibits an excellent fit to the data for a single 122-type phase with no evidence for any impuritiy phases.  The volume of the unit cell is 210.06(1)~\AA$^3$, which is $2.78\%$ larger than that of pure BaFe$_2$As$_2$, as expected for a single-phase sample.  These observations and those above in Sec.~\ref{SecSlowCooled} demonstrate that BaMn$_2$As$_2$ and BaFe$_2$As$_2$ are miscible at 1000\,$^\circ$C at the compositions Ba(Mn$_{0.8}$Fe$_{0.2}$)$_2$As$_2$ and Ba(Mn$_{0.2}$Fe$_{0.8}$)$_2$As$_2$ but phase-separate when slowly cooled from this temperature.

\begin{figure}
	\includegraphics[width=3in]{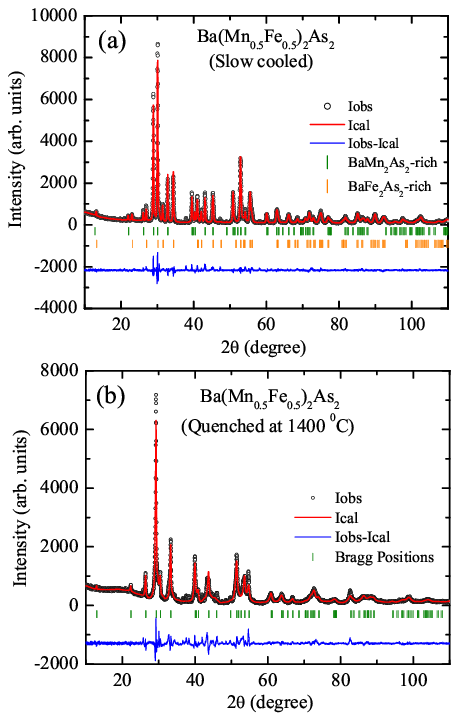}
	\caption{(Color online) Room-temperature powder XRD patterns (open circles) of (a) a slow-cooled sample of Ba(Mn$_{0.5}$Fe$_{0.5}$)$_2$As$_2$ and (b) a Ba(Mn$_{0.5}$Fe$_{0.5}$)$_2$As$_2$ sample quenched from 1400\,$^\circ$C to 77~K\@. The solid curve in (a) is a two-phase (BaMn$_2$As$_2$-rich and BaFe$_2$As$_2$-rich 122-type phases) Rietveld refinement fit whereas in (b) only a single 122-type phase is seen.  Difference profiles and Bragg peak positions of the fitted phases are also shown.  Bragg peaks from impurity phases  which are mainly Fe$_{1-x}$Mn$_x$As and FeAs$_2$ are also seen in the XRD patterns for both samples.}
	\label{fig:Figure_3_AP1017_XRD}
\end{figure}

In contrast to the samples with compositions $x = 0.2$ and 0.8 that were single-phase when quenched from 1000~$^\circ$C to 77 K, the samples with $x = 0.4$, 0.5 and~0.6 synthesized under identical conditions were manifestly multi-phase, demonstrating that the miscibility gap at 1000~$^\circ$C is  $0.2 \lesssim x \lesssim 0.8$.  The x-ray patterns of the latter three compositions were too ill-defined to be able to obtain reliable Rietveld refinements using them.  On increasing the temperature to 1100Ð-1400~$^\circ$C from which the samples were quenched (see Table~\ref{Table:CrystalData}), multi-phase samples were still obtained but for which the x-ray patterns could be refined assuming an approximately single 122-type phase plus impurity phases, suggesting that the miscibility gap does not close even at 1400~$^\circ$C\@.  For example, shown in Fig.~\ref{fig:Figure_3_AP1017_XRD} are powder XRD data for Ba(Mn$_{0.5}$Fe$_{0.5}$)$_2$As$_2$ samples that were either slow-cooled from 1000\,$^\circ$C or quenched from 1400\,$^\circ$C as representative of the \emph{x} =  0.4, 0.5 and 0.6 compositions.  Both samples contain impurity phases that are mainly FeAs$_2$ and Fe$_{1-x}$Mn$_x$As.  A two-phase Rietveld refinement of the XRD data for the slow-cooled sample in Fig.~\ref{fig:Figure_3_AP1017_XRD}(a) yields the unit cell volumes and percentage molar fractions of BaMn$_2$As$_2$-rich and BaFe$_2$As$_2$-rich phases shown in Table~I and Figs.~\ref{fig:Figure_Lattice_Parameters} and~\ref{fig:Figure_Molar_Fraction} that were discussed above in Sec.~\ref{SecSlowCooled}.  The powder XRD pattern of the quenched sample in Fig.~\ref{fig:Figure_3_AP1017_XRD}(b) indicates that the sample contains an approximately single 122-type phase in addition to impurity phases.  The calculated unit cell volume of the 122-type phase is 225.74(5)~\AA$^3$, which suggests a significant mixing of BaMn$_2$As$_2$ and BaFe$_2$As$_2$ at 1400\,$^\circ$C, but as noted this sample is not single-phase.  The crystal data for these quenched samples with compositions \emph{x} =  0.4, 0.5 and 0.6 are summarized in Table~\ref{Table:CrystalData} and Figs.~\ref{fig:Figure_Lattice_Parameters} and~\ref{fig:Figure_Molar_Fraction}.

The results in this section indicate that the miscibility gap $0.12 \lesssim x \lesssim 1$ at 300~K has narrowed to $0.2 \lesssim x \lesssim 0.8$ at 1000\,$^\circ$C\@.  The occurrence of a single 122-type phase plus impurity phases for samples with $0.2 \lesssim x \lesssim 0.8$ quenched from 1100 to 1400~$^\circ$C suggests the existence of a homogeneity range at these temperatures for the 122-type phase.

\subsection{\label{Sec:MnFeXtal_Compositions}  Ba(Mn$_x$Fe$_{1-x}$)$_2$As$_2$ Single Crystals}

\begin{figure}
	\includegraphics[width=2.5in]{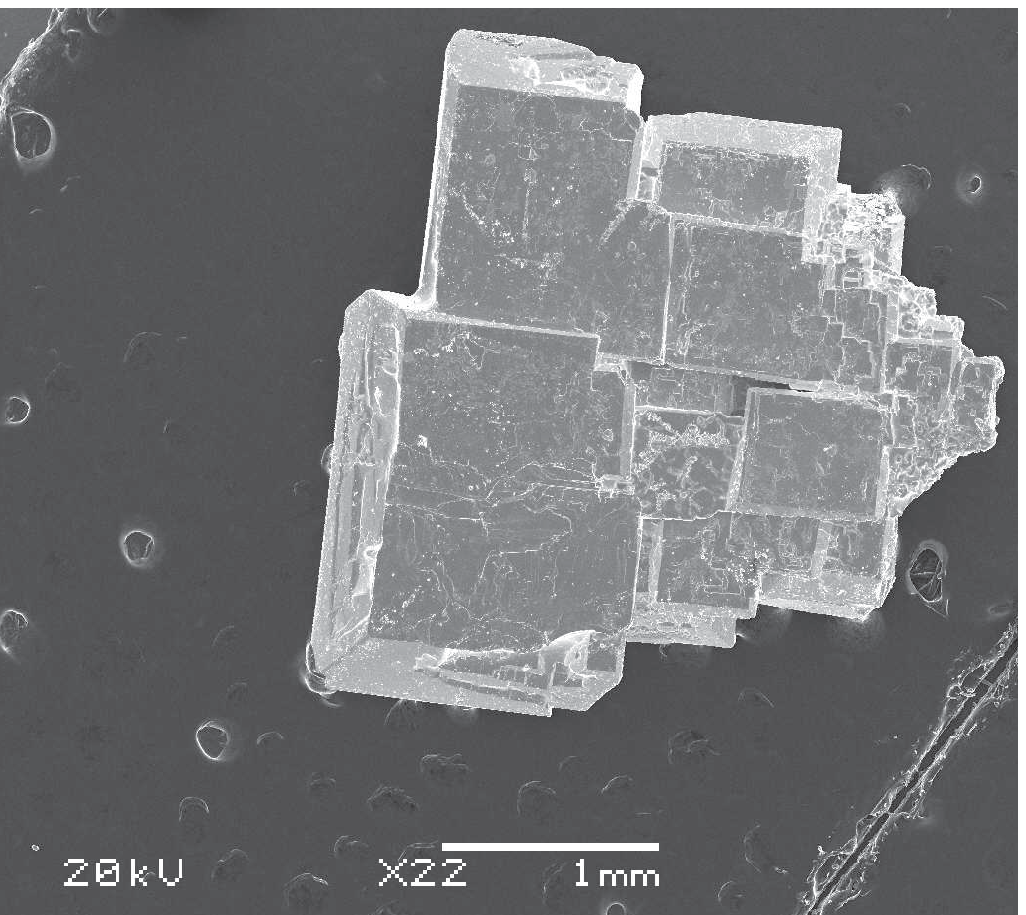}
	\includegraphics[width=2.5in]{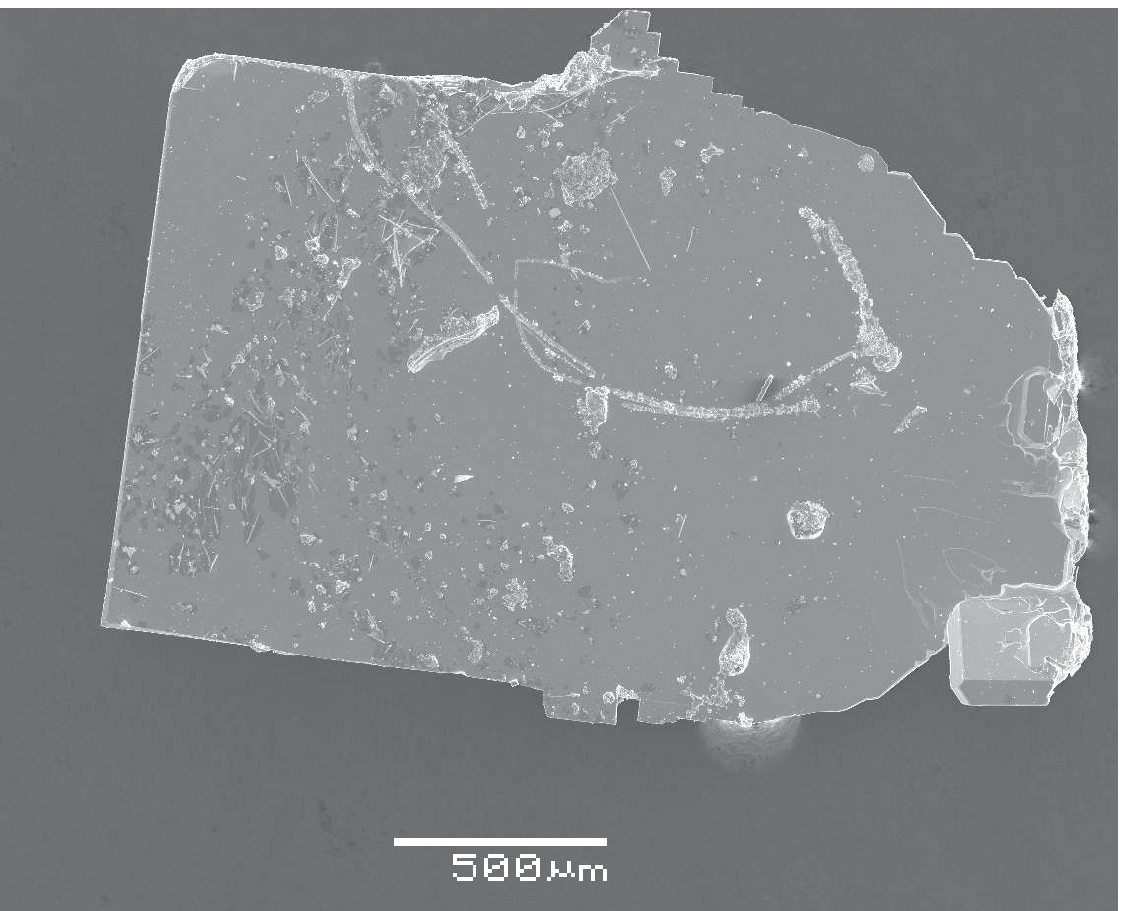}
	\caption{Scanning electron micrographs of Ba(Mn$_x$Fe$_{1-x}$)$_2$As$_2$ crystals grown in Sn flux from a starting composition of ${\rm Ba(Mn_{0.5}Fe_{0.5})_2As_2}$ upon slowly cooling from 1100\,$^\circ$C\@.  The top picture shows ``Crystal-2'' which contains intergrowths of crystals with compositions as listed for Crystal-2 in Table~\ref{Table:50:50 Mn-Fe Xtals} for which the decanting temperature was 700$\,^\circ$C\@.  The bottom picture is of a large single crystal ``Crystal-8'' (and a small crystal of the same composition attached to it on the bottom right) with the composition given for Crystal-8 in Table~\ref{Table:50:50 Mn-Fe Xtals}) for which the decanting temperature was 900$\,^\circ$C\@.  In both pictures, the large flat surfaces in the foreground are the $ab$~planes of the crystals.}
	\label{fig:AP1017_Crystal_1_SEM}
\end{figure}

\begin{table}
\caption{\label{Table:50:50 Mn-Fe Xtals}  Mn and Fe contents of Ba(Mn$_x$Fe$_{1-x}$)$_2$As$_2$ single crystals grown from a polycrystalline sample of nominal composition ${\rm Ba(Mn_{0.5}Fe_{0.5})_2As_2}$ using Sn flux that was decanted using a centrifuge at either 700 or 900~$^\circ$C, as listed, equivalent to a slow quench to room temperature from the respective temperature. The chemical composition was determined at up to six basal-plane spots on each crystal, as listed, using EDX analysis.  Of eleven crystals examined for which the Sn was decanted at $900\,^\circ$C (only five such crystals are shown in the table), none were found to be Fe-rich.  Note that ``Crystal-2'' is an intergrowth of Mn-rich and Fe-rich crystals shown in the top panel of Fig.~\ref{fig:AP1017_Crystal_1_SEM}.}

 \begin{ruledtabular}
		\begin{tabular}{l c c c }
		 Crystal &  Spot &  Mn-content & Fe-content \\
		         &       &    $x$ (\%)      & $100-x$ (\%) \\
		 \hline
			 Crystal-1 (700 $^\circ$C) & 1 & $8.7\pm 0.9$   & Remainder       \\
			                           & 2 & $8.4\pm 0.9$   & Remainder       \\
			 Crystal-2 (700 $^\circ$C) & 1 & $8.0\pm 0.9$   & Remainder       \\
			                           & 2 & Remainder      & $6.2\pm 0.6$   \\
			                           & 3 & Remainder      & $6.9\pm 0.6$   \\
			                           & 4 & $9\pm 1$       & Remainder       \\ 
			 Crystal-3 (700 $^\circ$C) & 1 & $10.4\pm 0.6$  & Remainder       \\
			                           & 2 & $8 \pm 1$      & Remainder       \\    
			                           & 3 & $10\pm 1$      & Remainder       \\ 
			                           & 4 & $9\pm 1$       & Remainder       \\ 
			                           & 5 & $16\pm 1$      & Remainder       \\
			                           & 6 & $15\pm 1$      & Remainder       \\ 
			 Crystal-4 (900 $^\circ$C) & 1 & Remainder      & $9.7 \pm 0.6$  \\
			                           & 2 & Remainder      & $10.5\pm 0.5$  \\
			                           & 3 & Remainder      & $9.7 \pm 0.5$  \\
			                           & 4 & Remainder      & $9.8 \pm 0.5$  \\
			 Crystal-5 (900 $^\circ$C) & 1 & Remainder      & $8.8 \pm 0.5$  \\
			                           & 2 & Remainder      & $9.2 \pm 0.5$  \\
			                           & 3 & Remainder      & $7.9 \pm 0.5$  \\
			 Crystal-6 (900 $^\circ$C) & 1 & Remainder      & $10.8 \pm 0.5$ \\
			                           & 2 & Remainder      & $10.4 \pm 0.9$ \\
			                           & 3 & Remainder      & $11.3 \pm 0.5$ \\  
       Crystal-7 (900 $^\circ$C) & 1 & Remainder      & $9.8 \pm 0.5$  \\
			                           & 2 & Remainder      & $9.1 \pm 0.5$  \\
			                           & 3 & Remainder      & $10.1 \pm 0.5$ \\ 
			 Crystal-8 (900 $^\circ$C) & 1 & Remainder      & $11.1 \pm 0.6$ \\
			                           & 2 & Remainder      & $10.2 \pm 0.9$ \\
			                           & 3 & Remainder      & $8.5  \pm 0.5$ \\			                                                  
		\end{tabular}
\end{ruledtabular}
\end{table}

Two batches of Ba(Mn$_x$Fe$_{1-x}$)$_2$As$_2$ crystals were grown in Sn flux from a polycrystalline sample of nominal composition ${\rm Ba(Mn_{0.5}Fe_{0.5})_2As_2}$ by slowly cooling from 1100\,$^\circ$C as discussed in Sec.~\ref{ExpDetails}.  The molten Sn flux was decanted at either 700 or 900\,$^\circ$C by removing the silica tube from the oven at the respective temperature and quickly placing the inverted tube into a centrifuge that spun up within $\sim 10$~s.  This procedure is equivalent to a slow quench of the crystals to room temperature from the respective temperature.  Examples of such crystals are shown in Fig.~\ref{fig:AP1017_Crystal_1_SEM}.  The compositions of many crystals from the two batches at up to six points on each crystal were measured using EDX and representative results are shown in Table~\ref{Table:50:50 Mn-Fe Xtals}.  The data for the crystals examined demonstrate that a miscibility gap exists in the Ba(Mn$_x$Fe$_{1-x}$)$_2$As$_2$ system, with a composition range of $0.16 \lesssim x \lesssim 0.93$ at 700~$^\circ$C and an upper limit to $x$ that decreases somewhat to $x_{\rm max} \approx 0.89$ at 900~$^\circ$C, qualitatively similar to our results above on polycrystalline samples.  Note that Crystal-3 obtained by decanting the Sn flux at 700~$^\circ$C showed a $\sim \pm 30$\% variation in the Mn concentration at different points on the crystal.

Among eleven crystals for which the Sn flux was decanted at $900\,^\circ$C that were examined by EDX analysis (only five of which are shown in Table~\ref{Table:50:50 Mn-Fe Xtals}), none were found to be Fe-rich even though such crystals were found when the Sn flux was decanted at $700\,^\circ$C\@.  This difference suggests that Fe-rich crystals grow in Sn flux only between 700 and $900\,^\circ$C when the starting composition is ${\rm Ba(Mn_{0.5}Fe_{0.5})_2As_2}$.

\section{\label{SecPowderProps} Physical Properties of Single-Phase Quenched Polycrystalline B\lowercase{a}(M\lowercase{n}$_{x}$F\lowercase{e}$_{1-x}$)$_2$A\lowercase{s}$_2$ Samples ($x$ = 0.2, 0.8)}

\subsection{Electrical Resistivity Measurements}

\begin{figure}
	\includegraphics[width=3in]{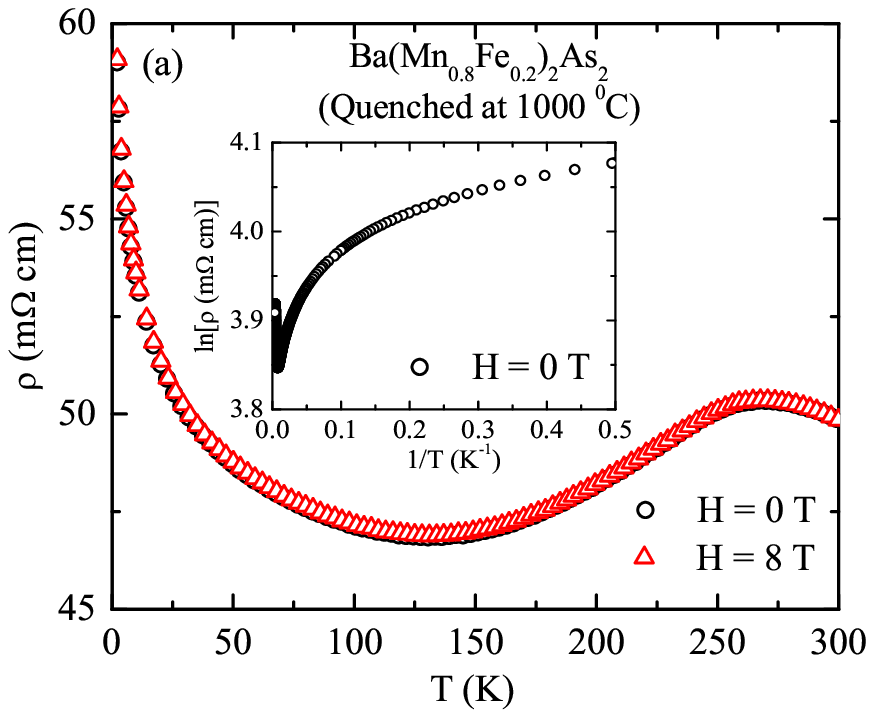}
	\includegraphics[width=3in]{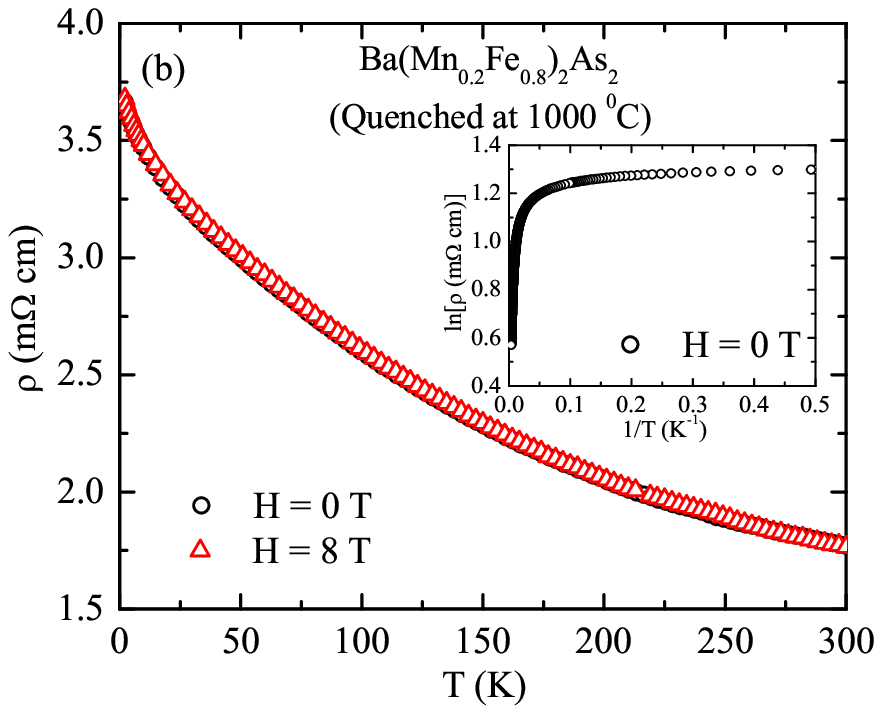}
	\caption{(Color online) Electrical resistivity $\rho$ of single-phase quenched polycrystalline samples of (a) Ba(Mn$_{0.8}$Fe$_{0.2}$)$_2$As$_2$ and (b) Ba(Mn$_{0.2}$Fe$_{0.8}$)$_2$As$_2$ versus temperature $T$ measured in applied magnetic fields $H = 0$ and 8~T\@. Insets: ln$\rho$ versus $1/T$.}
	\label{fig:Figure_7_Resistivity_AP1066}
\end{figure}

The $\rho(T)$ data for the single-phase quenched polycrystalline samples of Ba(Mn$_{0.8}$Fe$_{0.2}$)$_2$As$_2$ and Ba(Mn$_{0.2}$Fe$_{0.8}$)$_2$As$_2$ are shown in Fig.~\ref{fig:Figure_7_Resistivity_AP1066}. The large magnitudes and semiconducting-like temperature dependences suggest nonmetallic ground states for both compounds. However, we emphasize that electronic transport measurements on polycrystalline samples can give spurious results.  Measurements on single crystals are needed for definitive electronic transport characterizations.  The $\rho(T)$ data in applied magnetic fields $H = 0$ and $H = 8$~T closely superimpose, suggesting that spin-dependent scattering does not contribute significantly to the resistivity of these materials over the $T$ range of the measurements.  Plots of ln$\rho$ versus $1/T$ are shown in the insets of the respective figures. Extended linear regions are not found in these plots, which precludes the estimation of activation energies $\Delta$ in contrast to BaMn$_2$As$_2$ crystals where $\Delta \sim 30$~meV was found from in-plane $\rho(T)$ measurements.\cite{YSingh1}

The $\rho(T)$ data for Fe-rich ${\rm Ba(Mn_{0.2}Fe_{0.8})_2As_2}$ in Fig.~\ref{fig:Figure_7_Resistivity_AP1066}(b) show no obvious anomaly that might arise from a structural and/or antiferromagnetic (or spin density wave SDW) transition below 300~K\@.  We note that the results in Ref.~\onlinecite{Kim2010b} for single crystals of the Ba(Mn$_x$Fe$_{1-x}$)$_2$As$_2$ system with $0 \leq x \leq 0.176$ show evidence for stripe SDW order for $x = 0.176$ below a possibly smeared-out N\'eel temperature $T_{\rm N} \sim 200$~K [see Fig.~4(a) in Ref.~\onlinecite{Kim2010b}] and with a $T = 0$ ordered moment of $0.6\,\mu_{\rm B}$ per transition metal atom.  On the other hand, our $\rho(T)$ data in Fig.~\ref{fig:Figure_7_Resistivity_AP1066}(a) for Ba(Mn$_{0.8}$Fe$_{0.2}$)$_2$As$_2$, near the opposite Mn-rich end of the composition range, suggest that some type of phase transition may occur at $T \approx 260$~K\@.

\subsection{Magnetization and Magnetic Susceptibility Measurements}

\begin{figure}
	\includegraphics[width=3in]{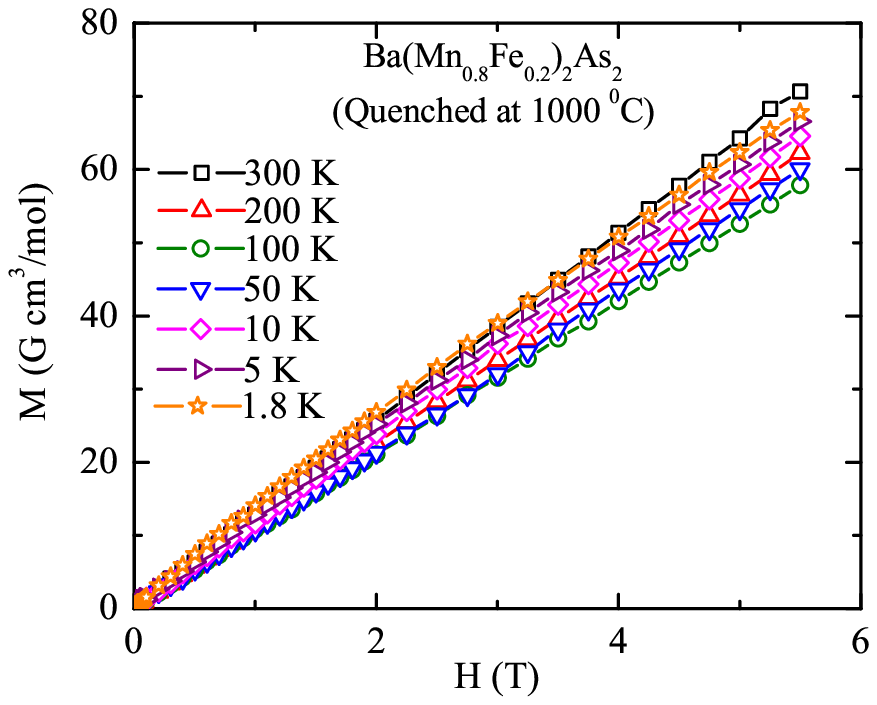}
	\includegraphics[width=3in]{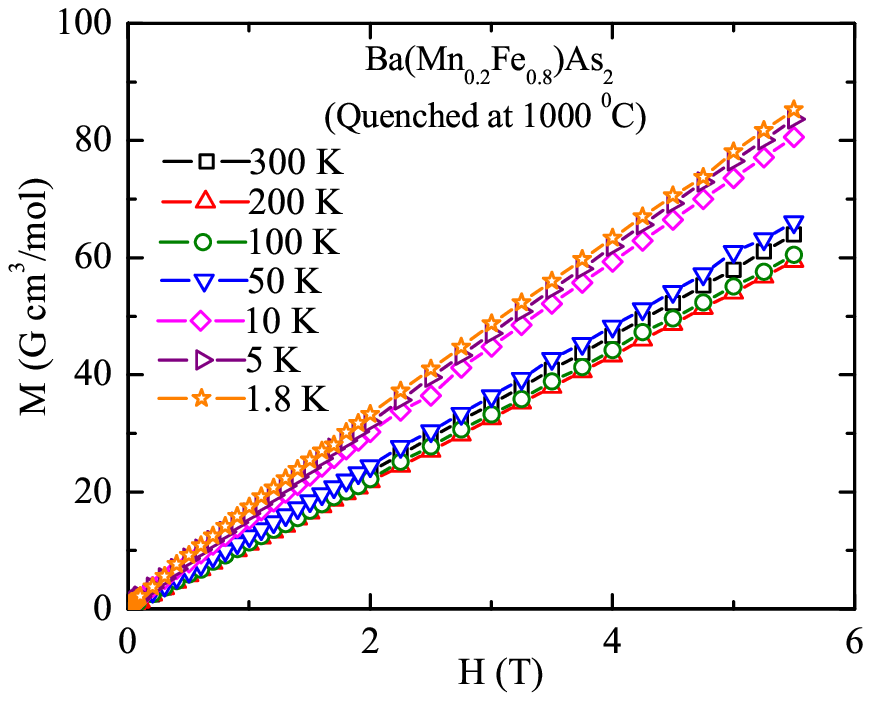}
	\caption{(Color online) Magnetization $M$ versus applied magnetic field $H$ isotherms at the indicated temperatures for single-phase polycrystalline Ba(Mn$_{0.8}$Fe$_{0.2}$)$_2$As$_2$ (top panel) and Ba(Mn$_{0.2}$Fe$_{0.8}$)$_2$As$_2$ (bottom panel). The solid lines are guides to the eye.}
	\label{fig:Figure_5_MH_AP1066}
\end{figure}

\begin{figure}
	\includegraphics[width=3in]{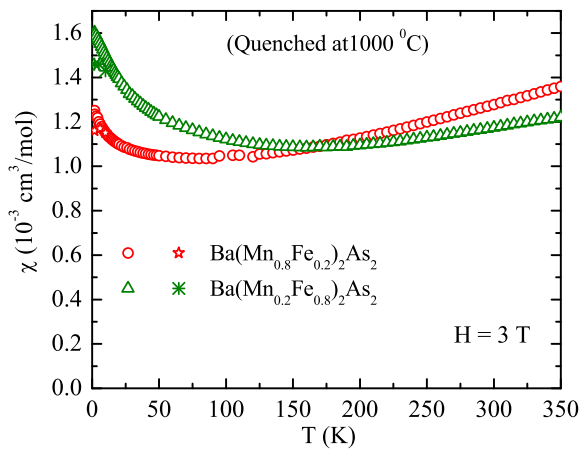}
	\caption{(Color online) Magnetic susceptibility $\chi \equiv M/H$ versus temperature $T$ of single-phase quenched polycrystalline Ba(Mn$_{0.8}$Fe$_{0.2}$)$_2$As$_2$ and Ba(Mn$_{0.2}$Fe$_{0.8}$)$_2$As$_2$ samples measured in an applied magnetic field $H = 3$~T\@. Stars and asterisks represent the low-temperature $\chi$ values extracted at $T \leq 10$~K from $M(H)$ isotherms for these two compounds, respectively.}
	\label{fig:Figure_4_MT_All}
\end{figure}

Isothermal magnetization $M$ versus $H$ measurements for the quenched single-phase polycrystalline samples of Ba(Mn$_{0.8}$Fe$_{0.2}$)$_2$As$_2$ and Ba(Mn$_{0.2}$Fe$_{0.8}$)$_2$As$_2$ are shown in Fig.~\ref{fig:Figure_5_MH_AP1066} for the field range up to 5.5~T\@.  The $M$ data taken at various temperatures for both compositions are proportional to $H$, except for the data below 10~K where they exhibit slight nonlinear behaviors with negative curvature for fields below $\sim 2$~T, which most likely arise from saturable paramagnetic impurities.  We performed linear fits to the $M(H)$ data in the field range $3\le H \le 5.5$~T for $T \leq 10$~K; the high-field slopes give independent values of $\chi(T)$ of the materials below 10~K that are referred to below. 

The magnetic susceptibilities $\chi\equiv M/H$ versus temperature $T$ measured in $H = 3$~T for the quenched samples of Ba(Mn$_{0.8}$Fe$_{0.2}$)$_2$As$_2$ and Ba(Mn$_{0.2}$Fe$_{0.8}$)$_2$As$_2$ are shown in Fig.~\ref{fig:Figure_4_MT_All}, together with the values of the $\chi(T)$ calculated as described above from the $M(H)$ isotherms for $T\le 10$~K\@. These latter values are similar to the values obtained in the respective $M(T)$ scans at fixed $H$, demonstrating that the amounts of saturable impurities are almost negligible in these samples.  The positive slopes of $\chi(T)$ for Ba(Mn$_{0.8}$Fe$_{0.2}$)$_2$As$_2$ and Ba(Mn$_{0.2}$Fe$_{0.8}$)$_2$As$_2$ at the higher temperatures are consistent with the similar behaviors reported in the paramagnetic states for single crystals of BaMn$_2$As$_2$ (Ref.~\onlinecite{YSingh1}) and BaFe$_2$As$_2$.\cite{Wang2009, Rotter2} The upturns in $\chi(T)$ for the samples below $\sim 70$--200~K may be due to the presence of nonsaturable paramagnetic impurities and/or magnetic defects, although as previously noted we did not detect any impurity phases in these samples from XRD measurements. There is no clear evidence from the $\chi(T)$ data for either sample for an anomaly in $\chi(T)$ due a SDW and/or tetragonal-to-orthorhombic transition.  Such transitions are observed at $\approx 137$~K in undoped BaFe$_2$As$_2$.\cite{Wang2009, Rotter2}

We note that consistent with the XRD data in Fig.~\ref{fig:Figure_3_AP1017_XRD}(b) for the quenched (1400\,$^\circ$C to 77~K) sample of Ba(Mn$_{0.5}$Fe$_{0.5}$)$_2$As$_2$, our $M(H,T)$ data for this sample (not shown) indicated the presence of a substantial amount of ferromagnetic impurities.  Hence, we did not attempt to extract $\chi(T)$ from the $M(H,T)$ data for this sample.

\subsection{Heat Capacity Measurements}

\begin{figure}
	\includegraphics[width=3.3in]{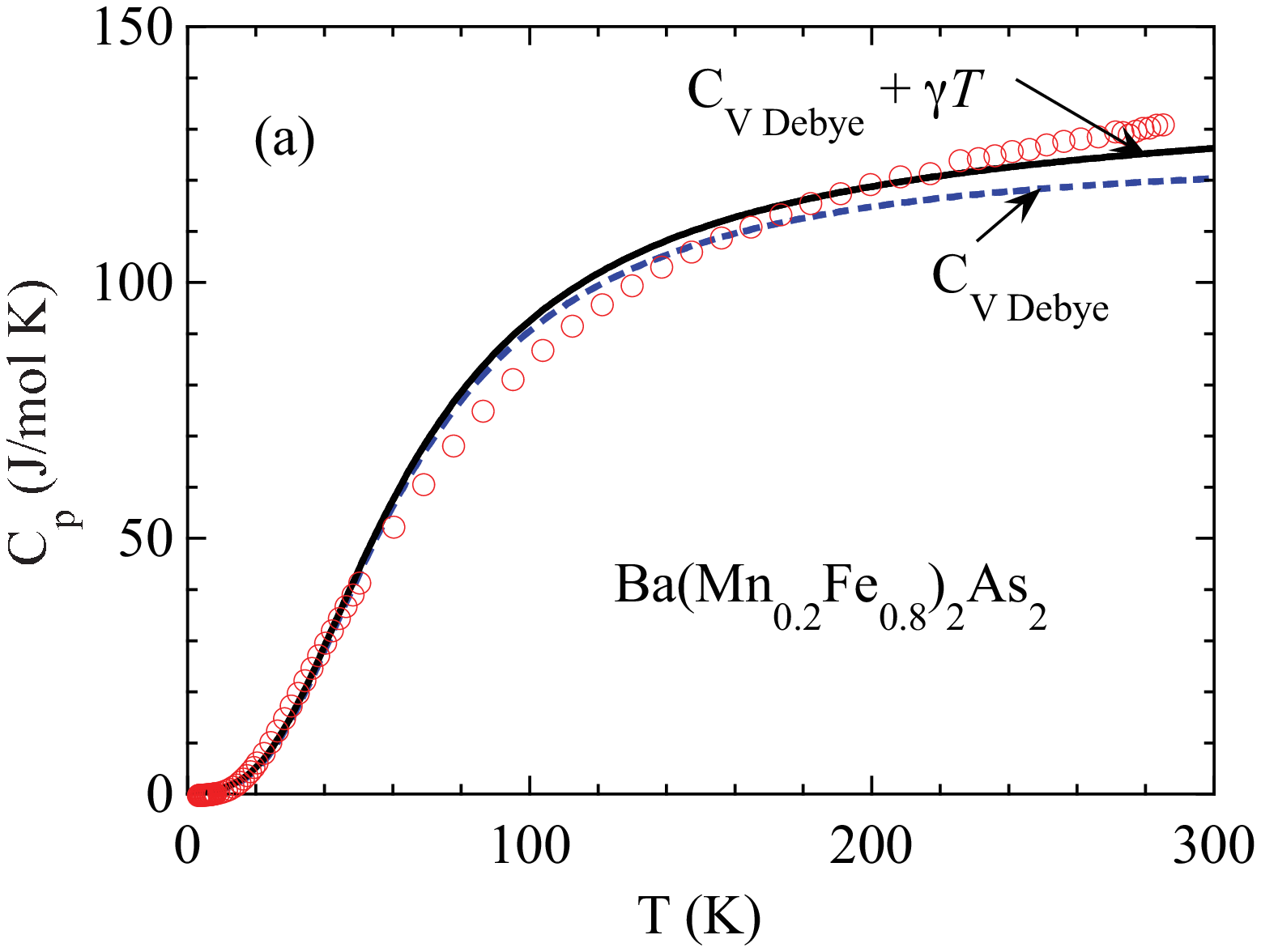}
	\includegraphics[width=3.3in]{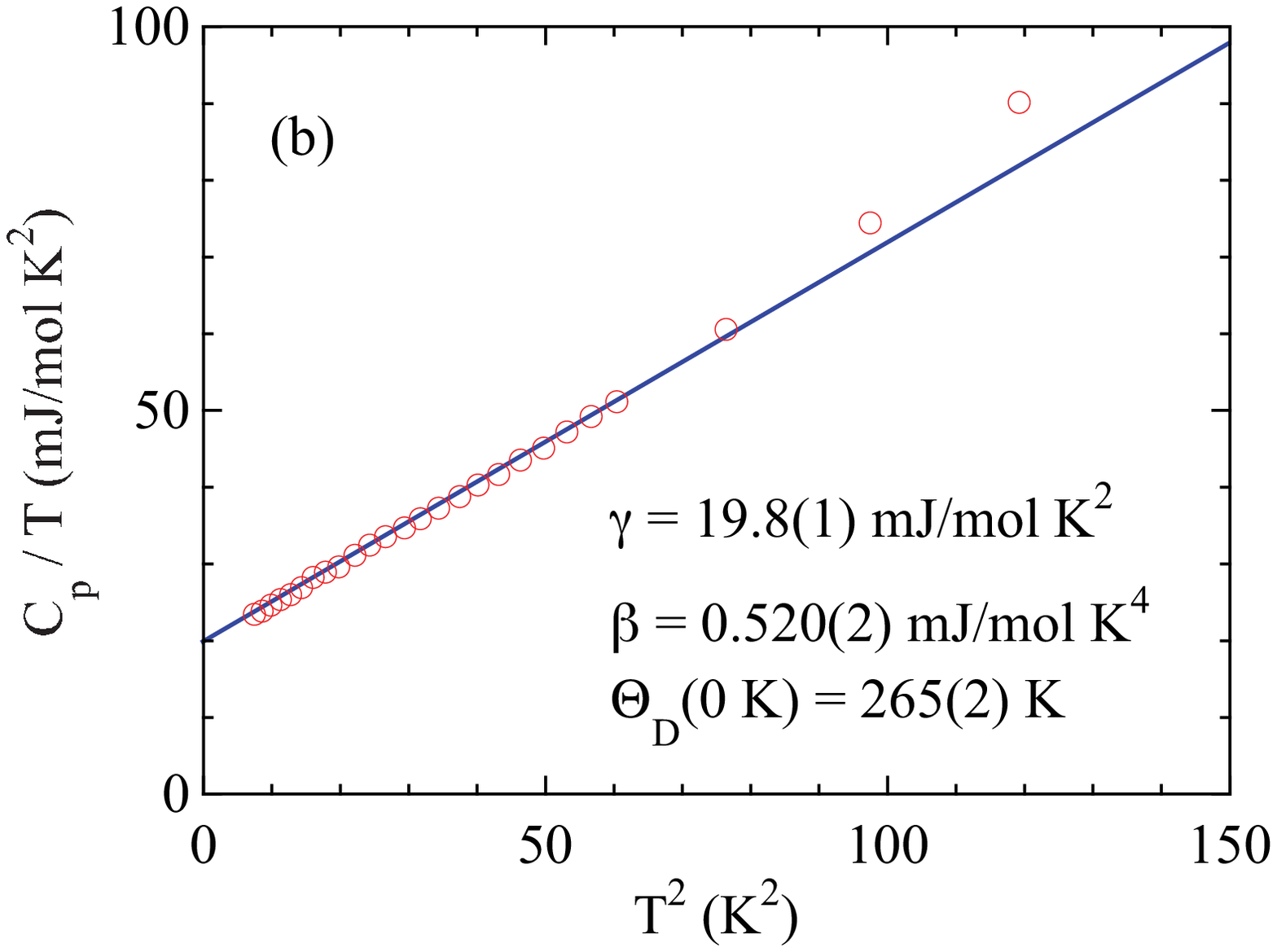}
	\caption{(Color online) (a) Heat capacity $C_{\rm p}$ versus temperature $T$ (open circles) for a single-phase polcrystalline sample of $\rm{Ba(Mn_{0.2}Fe_{0.8})_2As_2}$ quenched from 1000~$^\circ$C\@.  Also shown are the Debye lattice heat capacity function $C_{\rm V\ Debye}(T)$ obtained from Eq.~(\ref{Eq:CDebyeFcn}), and $C_{\rm V\ Debye}(T) + \gamma T$, where the parameters are the Debye temperature $\Theta_{\rm D} = 265$~K and $\gamma$ value determined from the low-temperature fit by Eq.~(\ref{Equation1}) in (b).}
	\label{fig:Ba(Mn0.2Fe0.8)2As2_Cp}
\end{figure}

\begin{figure}
	\includegraphics[width=3.3in]{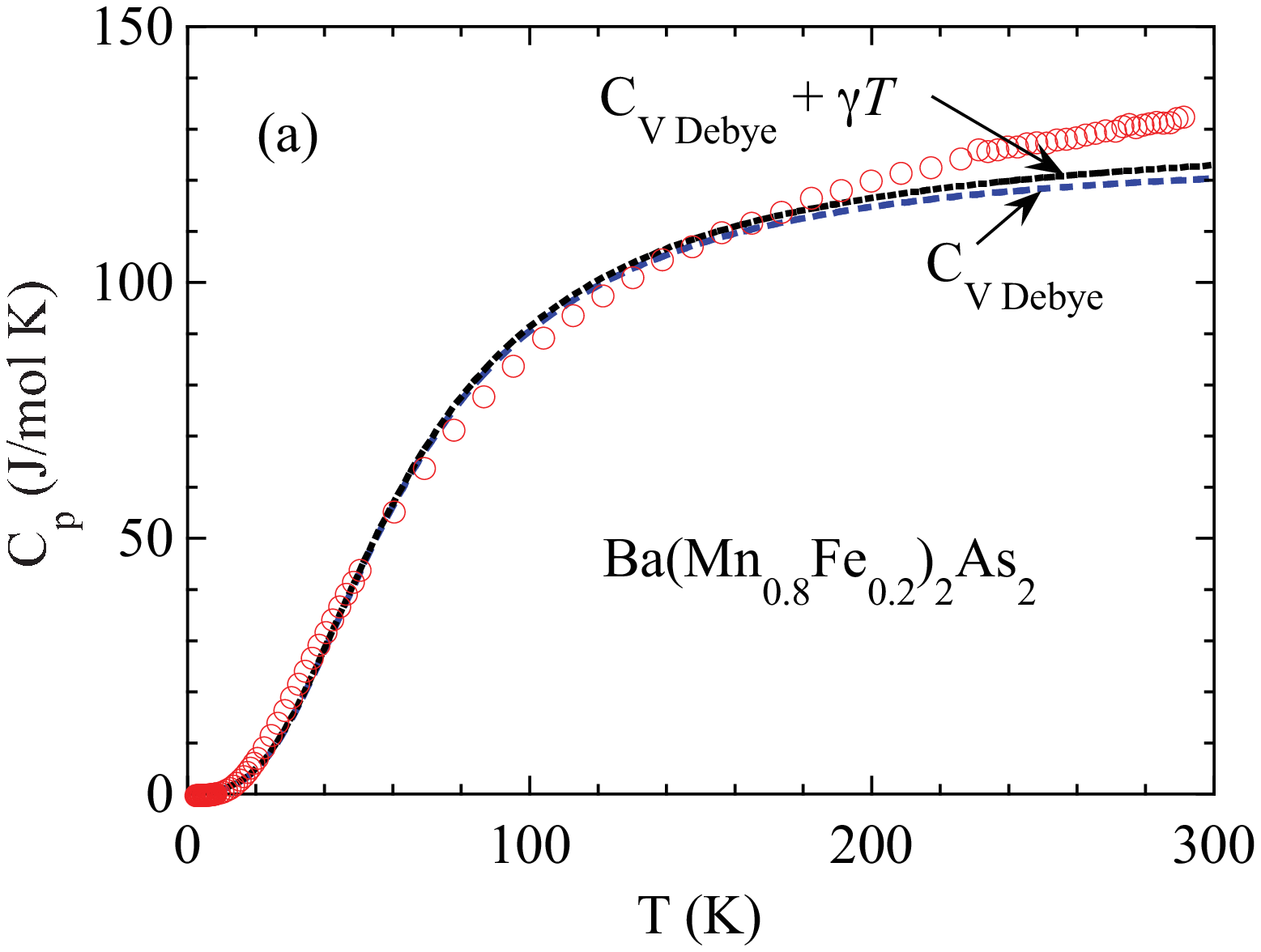}
	\includegraphics[width=3.3in]{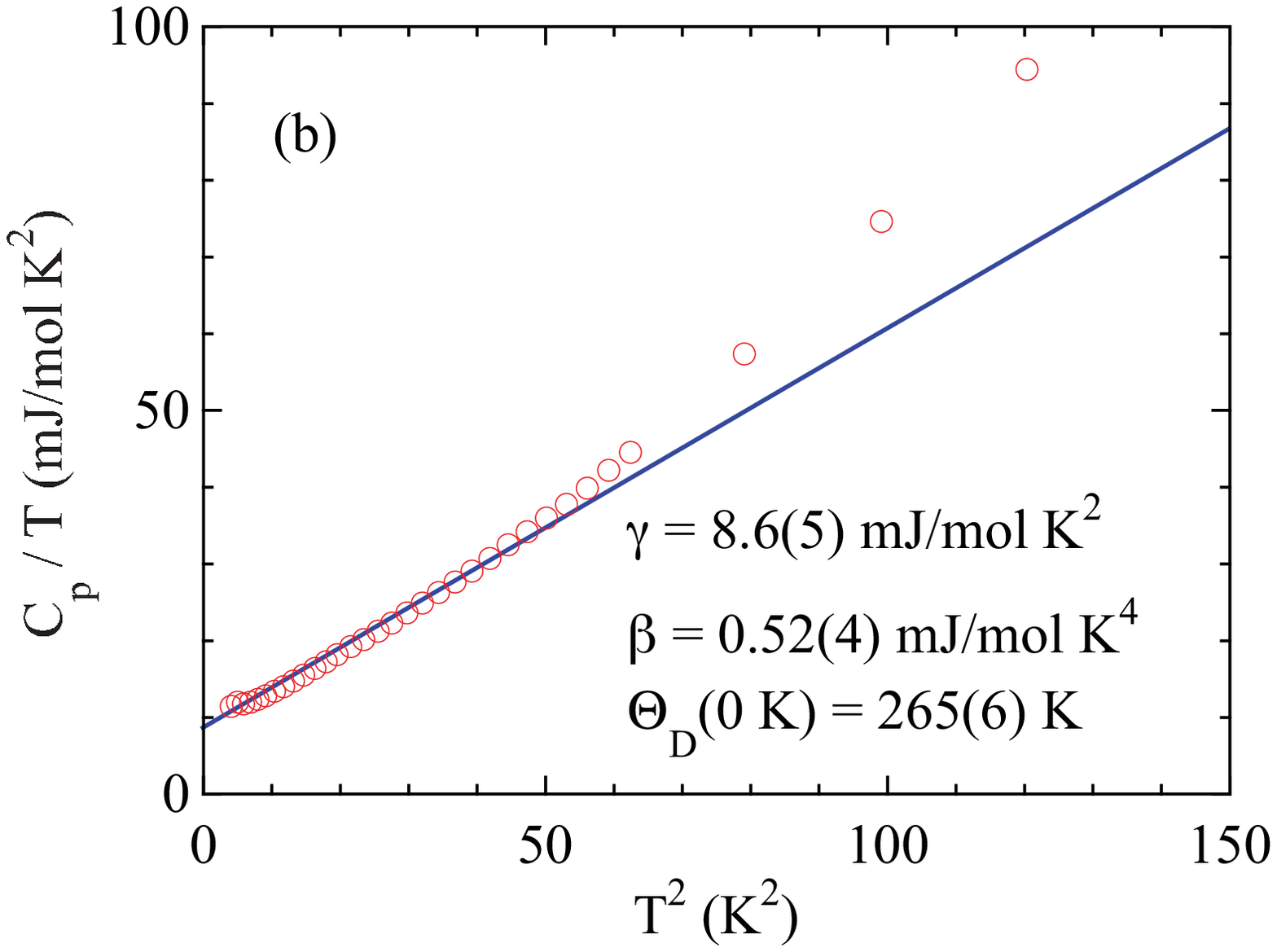}
	\caption{(Color online) The same types of heat capacity plots for $\rm{Ba(Mn_{0.8}Fe_{0.2})_2As_2}$ quenched from 1000~$^\circ$C as for $\rm{Ba(Mn_{0.2}Fe_{0.8})_2As_2}$ in Fig.~\ref{fig:Ba(Mn0.2Fe0.8)2As2_Cp}.}
	\label{fig:Ba(Mn0.8Fe0.2)2As2_Cp}
\end{figure}
The temperature variation of the heat capacity at constant pressure ($C_{\rm p}$) for the two single-phase quenched polycrystalline samples of $\rm{Ba(Mn_{0.2}Fe_{0.8})_2As_2}$ and $\rm{Ba(Mn_{0.8}Fe_{0.2})_2As_2}$ are shown in Figs.~\ref{fig:Ba(Mn0.2Fe0.8)2As2_Cp} and~\ref{fig:Ba(Mn0.8Fe0.2)2As2_Cp}, respectively.  The measured value of $C_{\rm p}$ at $T = 275$~K is 129 and 130 J/mol K for $\rm{Ba(Mn_{0.2}Fe_{0.8})_2As_2}$ and $\rm{Ba(Mn_{0.8}Fe_{0.2})_2As_2}$, respectively. These values are close to, but a bit larger than, the classical Dulong-Petit heat capacity $C_{\rm V} = 15R \approx 124.7$ J/mol K expected for a crystalline compound containing five atoms/f.u. The $C_{\rm p}(T)$ data for the two compounds show no evidence for any phase transitions in the temperature range of measurement.

We fitted the low-temperature $C_{\rm p}(T)$ data in Figs.~\ref{fig:Ba(Mn0.2Fe0.8)2As2_Cp}(b) and~\ref{fig:Ba(Mn0.8Fe0.2)2As2_Cp}(b) using the expression 
\begin{equation}
\frac{C_{\rm p}}{T} = \gamma+\beta T^2
\label{Equation1}
\end{equation}
where $\gamma$ is the Sommerfeld coefficient of the electronic heat capacity and $\beta$ is the  coefficient of the Debye $T^3$ term for lattice heat capacity. The values of the fitting parameters and their error bars are given in the figures. The value of $\gamma$ reported by Singh et al.\ for $\rm{BaMn_2As_2}$ is 0.0(4) mJ/mol K$^2$.\cite{YSingh1} The nonzero values of $\gamma$ for $\rm{Ba(Mn_{0.8}Fe_{0.2})_2As_2}$ and $\rm{Ba(Mn_{0.2}Fe_{0.8})_2As_2}$ suggest metallic ground states.  Indeed, our $\rho(T)$ measurements indicate that the magnitude of the resistivities of $\rm{Ba(Mn_{0.8}Fe_{0.2})_2As_2}$ and $\rm{Ba(Mn_{0.2}Fe_{0.8})_2As_2}$ are significantly smaller than reported for $\rm{BaMn_2As_2}$.\cite{YSingh1} The Debye temperature $\Theta_{\rm D}$ is estimated using the expression\cite{Kittel2005}
\begin{equation}
\Theta_{\rm D} = \left(\frac{12\pi^4Rn}{5\beta}\right)^{1/3}
\label{Equation2}
\end{equation}
where $n$ is the number of atoms per formula unit [$n=5$ in the present case of Ba(Mn$_{x}$Fe$_{1-x}$)$_2$As$_2$] and $R$ is the molar gas constant. The calculated values of $\Theta_{\rm D}$ are about 265~K, somewhat larger than $\Theta_{\rm D} = 246(4)$ K reported for a single crystal of $\rm{BaMn_2As_2}$.\cite{YSingh1}

The Debye lattice heat capacity function at constant volume per mole of atoms is\cite{Kittel2005}
\be
C_{\rm Debye\ V} = 9R\left(\frac{T}{\Theta_{\rm D}}\right)^3\int_0^{\Theta_{\rm D}/T}\frac{x^4e^x}{(e^x -1 )^2}dx.
\label{Eq:CDebyeFcn}
\ee
This function describes the heat capacity of acoustic phonons.  The $C_{\rm Debye\ V}(T)$ is plotted in Figs.~\ref{fig:Ba(Mn0.2Fe0.8)2As2_Cp}(a) and~\ref{fig:Ba(Mn0.8Fe0.2)2As2_Cp}(a) using the value $\Theta_{\rm D} = 265$~K determined in Figs.~\ref{fig:Ba(Mn0.2Fe0.8)2As2_Cp}(b) and~\ref{fig:Ba(Mn0.8Fe0.2)2As2_Cp}(b) and taking into account the five atoms per formula unit.  Overall, the Debye function fits the data reasonably well, but a temperature dependence of $\Theta_{\rm D}$ is evident from the comparison.\cite{Gopal1966}  Also, the experimental data at the higher temperatures are above the Debye function, suggesting additional contributions.  We added the respective electronic contributions $\gamma T$ to the Debye function, as shown in Figs.~\ref{fig:Ba(Mn0.2Fe0.8)2As2_Cp}(a) and~\ref{fig:Ba(Mn0.8Fe0.2)2As2_Cp}(a).  The data are still above the calculated values at the highest temperatures.  Finally, we calculated the difference  $C_{\rm p}-C_{\rm V}$ for the lattice heat capacity for the compound ${\rm BaFe_2As_2}$ according to the thermodynamic relation
\[
C_{\rm p}-C_{\rm V} = V_{\rm M}\beta_{\rm V}^2(T)B(T)T,
\]
where $V_{\rm M}$ is the molar volume, $\beta_{\rm V}$ is the volume thermal expansion coefficient, and $B$ is the bulk modulus.  For the 200--300~K temperature range, using the values $\beta_{\rm V} \approx 4.8\times10^{-5}$~K$^{-1}$,\cite{Budko2010} $B = 66\times10^{11}$~dyne/cm$^2$,\cite{Mittal2011} and $V_{\rm M}= 61.5$~cm$^3$/mol,\cite{Johnston2010} we obtained
\[
C_{\rm p}-C_{\rm V} \approx {\rm \left(9.3\,\frac{mJ}{mol\,K^2}\right)}T. \hspace{0.2in} ({\rm 200\ to\ 300~ K})
\]
This prediction is similar in magnitude to the electronic $\gamma T$ contribution for the Mn-rich sample in Fig.~\ref{fig:Ba(Mn0.8Fe0.2)2As2_Cp}(a) and is not sufficient to obtain agreement with both the magnitude and slope of the data near 300~K, especially for the Mn-rich sample, suggesting that at least one other contribution to the heat capacity is present.  This additional heat capacity could arise from optic phonons and/or from  magnetic excitations. 

\section{\label{SecXtalResults} Physical Properties of Single Crystalline C\lowercase{r}-, F\lowercase{e}- and S\lowercase{b}-substituted B\lowercase{a}M\lowercase{n}$_2$A\lowercase{s}$_2$}

As noted previously in Sec.~\ref{ExpDetails}, our attempts to grow single crystals of substituted Ba(Mn$_{1-x}T_x$)$_2$As$_2$ ($T$ = Cr, Fe, Co, Ni, Cu, Ru, Rh, Pd, Re, Pt) and BaMn$_2$(As$_{1-x}$Sb$_x$)$_2$ using Sn flux were mostly unsuccessful. The only partially successful attempts were with Cr and Fe substitutions for Mn and Sb substitution for As, where we achieved 4.4\%, $\sim 10$\% and 2.6\% substitutions, respectively. These concentrations were much smaller than present in the starting compositions of the crystal growth mixtures. Here we present in-plane $\rho(T)$ and anisotropic $\chi(T)$ measurements on such Cr-, Fe- and Sb-substituted crystals, which were found to be similar to the respective properties of unsubstituted BaMn$_2$As$_2$ crystals.

\subsection{Electrical Resistivity Measurements}

\begin{figure}
	\includegraphics[width=3.3in]{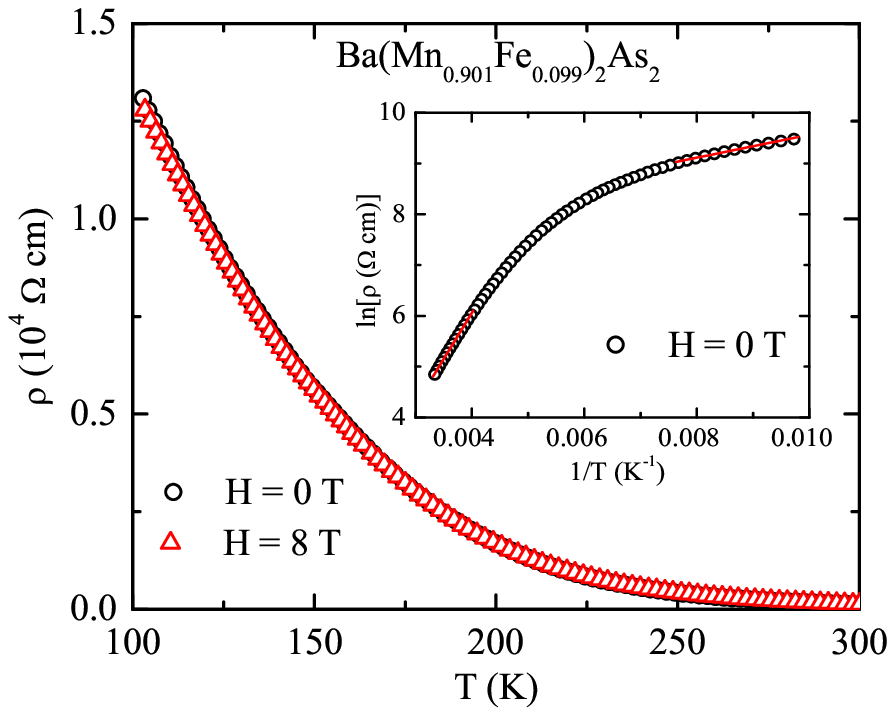}
	\includegraphics[width=3.3in]{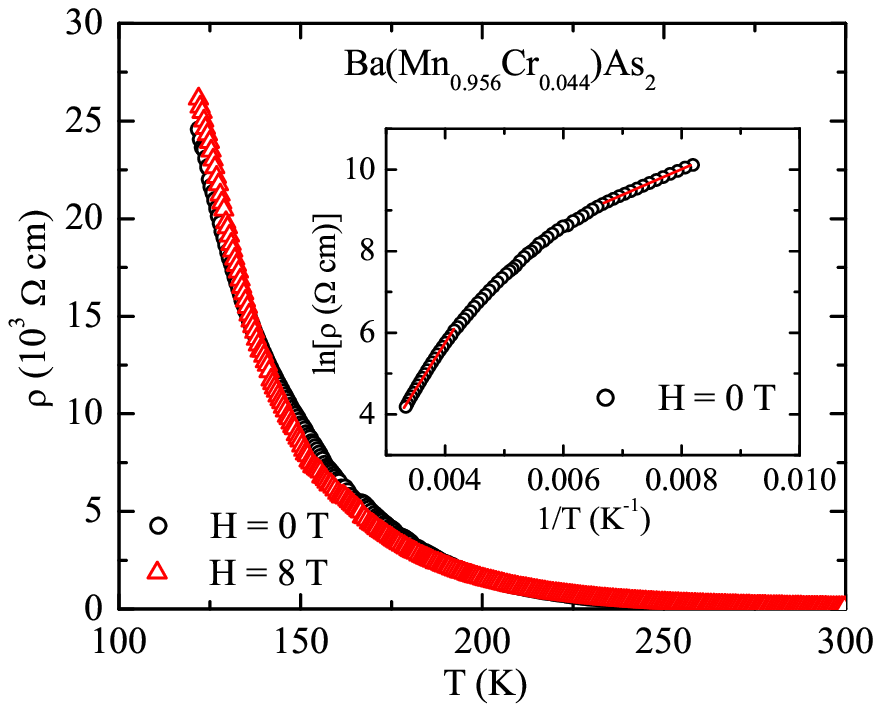}
	\caption{(Color online) In-plane electrical resistivity $\rho$ versus temperature $T$ of single crystals of ${\rm Ba(Mn_{0.901}Fe_{0.099})_2As_2}$ (top panel) and ${\rm Ba(Mn_{0.956}Cr_{0.044})_2As_2}$ (bottom panel) in applied magnetic fields $H=0$ and 8~T\@. Inset of each panel: ln$\rho$ versus $1/T$ at $H=0$~T\@. The two solid lines in the upper inset are fits of the data by the expression $\ln\rho=A+\Delta/T$ over the temperature ranges 100--130 and 250--300~K, which yield activation energies $\Delta = 19$ and~156~meV, respectively.  In the lower inset, the data were fitted from 120--150 and 250--300~K, as shown, yielding $\Delta = 54$ and~197~meV, respectively.}
	\label{fig:Figure_10_Resistivity_Crystal}
\end{figure}

The in-plane $\rho(T)$ data for single crystals of ${\rm Ba(Mn_{0.901}Fe_{0.099})_2As_2}$ and ${\rm Ba(Mn_{0.956}Cr_{0.044})_2As_2}$ are shown in Fig.~\ref{fig:Figure_10_Resistivity_Crystal}.  The $\rho(T)$ behaviors for both crystals exhibit monotonic increases with decreasing $T$ below our high-temperature limit of 300~K\@.  The resistance values below $\sim 100$~K became very large and exceeded the measurement limit of our instrument, indicating insulating ground states of the materials, similar to previous observations on single-crystal BaMn$_2$As$_2$.\cite{YSingh1, An1}  An applied magnetic field of 8~T resulted in no significant change in $\rho(T)$ for either sample, as shown.

The insets of the two panels in Fig.~\ref{fig:Figure_10_Resistivity_Crystal} show ln$\rho$ versus $1/T$.  The data in each inset suggest two quasi-linear regions, as shown by the two straight red lines. The data in these regions follow $\ln\rho=A+\Delta/k_{\rm B}T$, where $A$ is a constant, $\Delta$ is the activation energy and $k_{\rm B}$ is Boltzmann's constant. The fitted values of the activation energies in the lowest and highest-$T$ regions for each sample are given in the figure caption.  The larger value for each sample can probably be interpreted as the intrinsic activation energy while the smaller one may be the energy gap between donor or acceptor energy levels and a conduction or valence band, respectively.\cite{Pearson1}  These observations suggest that similar to BaMn$_2$As$_2$,\cite{YSingh1} ${\rm Ba(Mn_{0.901}Fe_{0.099})_2As_2}$ and ${\rm Ba(Mn_{0.956}Cr_{0.044})_2As_2}$ are small-band-gap semiconductors.

\subsection{Magnetization and Magnetic Susceptibility Measurements}
\begin{figure}
	\includegraphics[width=3in]{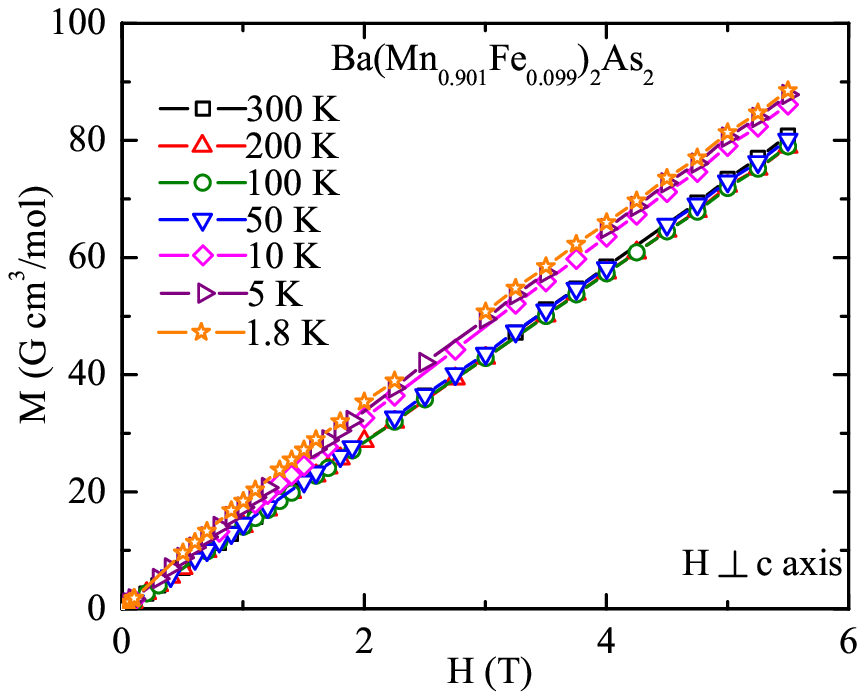}
	\includegraphics[width=3.1in]{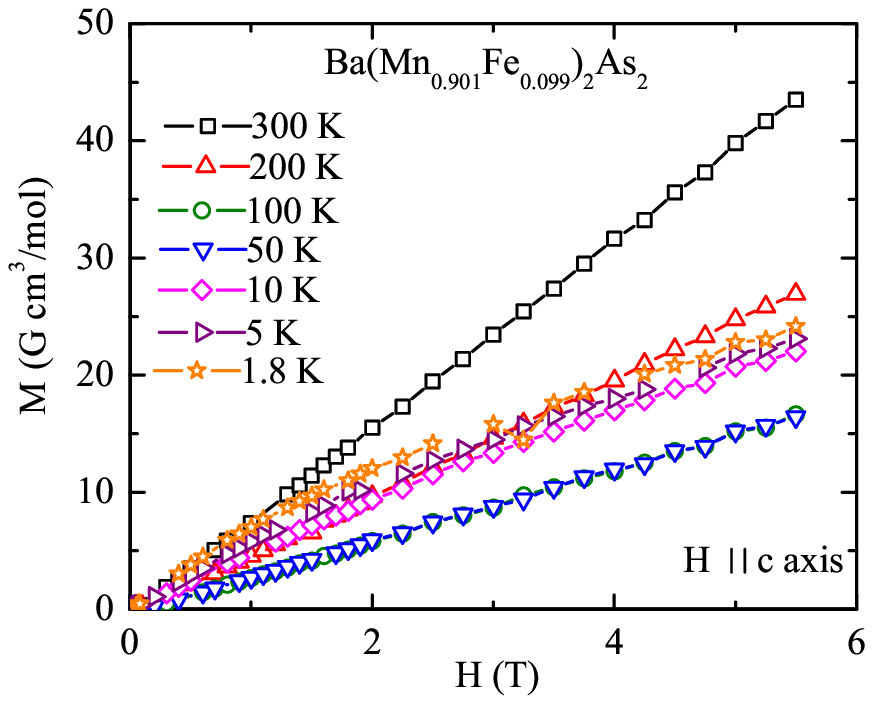}
	\caption{(Color online) Isothermal magnetization $M$ versus applied magnetic field $H$ in the $ab$~plane (top panel) and along the $c$~axis (bottom panel) for a single crystal of ${\rm Ba(Mn_{0.901}Fe_{0.099})_2As_2}$.}
	\label{fig:Figure_15_Fe-doped_BaMn2As2_MH}
\end{figure}

\begin{figure}
	\includegraphics[width=3in]{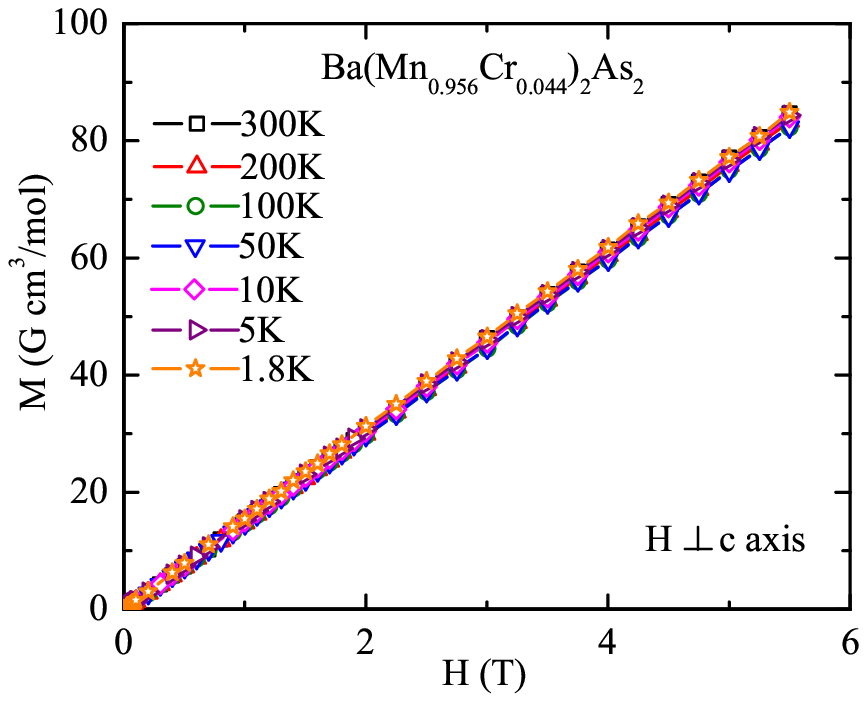}
	\includegraphics[width=3in]{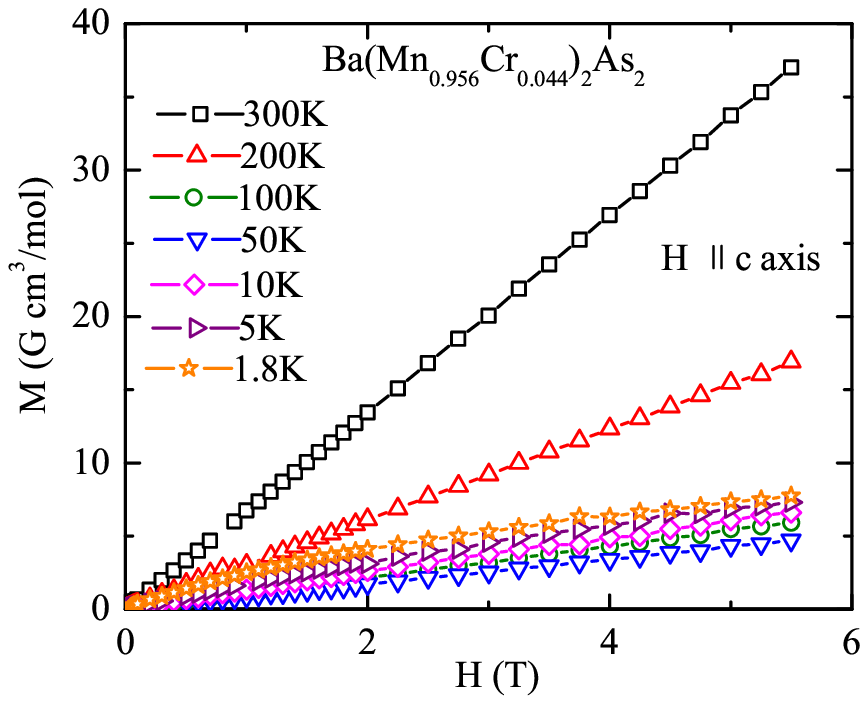}
	\caption{(Color online) Anisotropic magnetization $M$ versus applied magnetic field $H$ isotherms at the indicated temperatures for ${\rm Ba(Mn_{0.956}Cr_{0.044})_2As_2}$.  The magnetic field was applied either in the $ab$~plane (top panel) or along the $c$~axis (bottom panel).}
	\label{fig:Figure_Cr-doped-BaMn2As2_MH}
\end{figure}

\begin{figure}
	\includegraphics[width=3in]{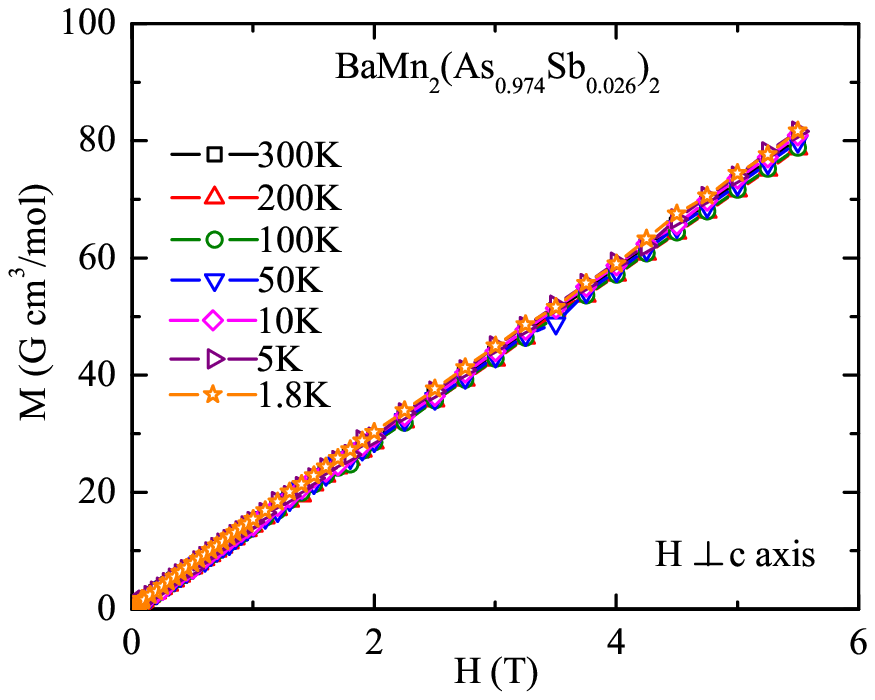}
	\includegraphics[width=3in]{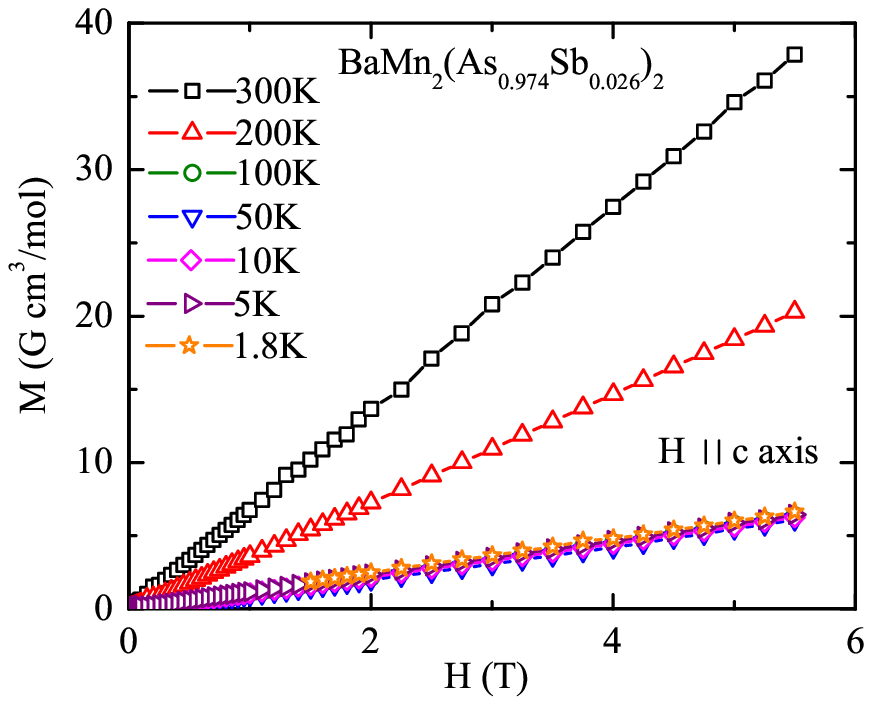}
	\caption{(Color online) Anisotropic magnetization $M$ versus applied magnetic field $H$ isotherms at the indicated temperatures for ${\rm BaMn_2(As_{0.974}Sb_{0.026})_2}$.  The magnetic field was applied either in the $ab$~plane (top panel) or along the $c$~axis (bottom panel).}
	\label{fig:Figure_18_Sb-doped-BaMn2As2_MH}
\end{figure}

Shown in Figs.~\ref{fig:Figure_15_Fe-doped_BaMn2As2_MH}, \ref{fig:Figure_Cr-doped-BaMn2As2_MH} and~\ref{fig:Figure_18_Sb-doped-BaMn2As2_MH} are $M(H)$ isotherms at temperatures from 1.8 to 300~K for single crystals of Ba(Mn$_{0.901}$Fe$_{0.099}$)$_2$As$_2$, ${\rm Ba(Mn_{0.956}Cr_{0.044})_2As_2}$ and ${\rm BaMn_2(As_{0.974}Sb_{0.026})_2}$, respectively.  For each compound $M$ is proportional to $H$, except for temperatures $T \le 10$~K where the data sometimes exhibit nonlinear behavior at fields below $\sim 2$~T\@.  These data are similar to the above-discussed $M(H,T)$ data for quenched polycrystalline samples of Ba(Mn$_{0.8}$Fe$_{0.2}$)$_2$As$_2$ and Ba(Mn$_{0.2}$Fe$_{0.8}$)$_2$As$_2$, and indicate the absence of significant amounts of ferromagnetic or saturable paramagnetic impurities in the samples.  In particular, the problematic relatively large amounts of ferromagnetic MnAs impurities present in BaMn$_2$As$_2$ crystals studied previously\cite{YSingh1} are not present in the present lightly substituted crystals.

\begin{figure}
	\includegraphics[width=\columnwidth]{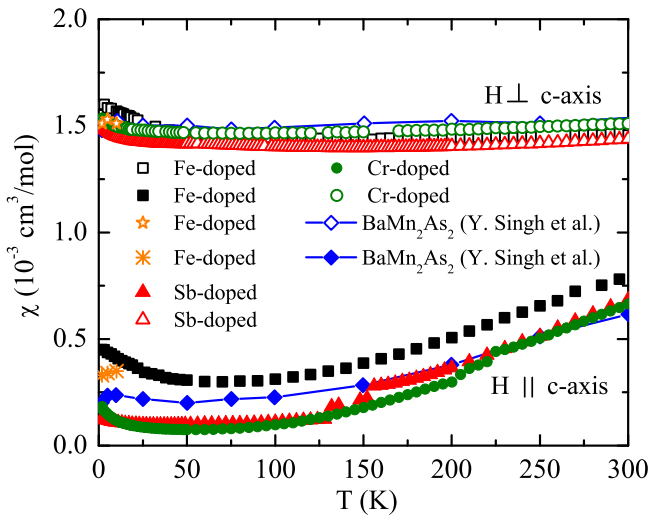}
	\caption{(Color online) Magnetic susceptibilities $\chi_{ab}$ and $\chi_{c}$ in the $ab$~plane and along the $c$~axis, respectively, of single crystals of Ba(Mn$_{0.901}$Fe$_{0.099}$)$_2$As$_2$, ${\rm Ba(Mn_{0.956}Cr_{0.044})_2As_2}$ and ${\rm BaMn_2(As_{0.974}Sb_{0.026})_2}$ versus temperature in an applied magnetic field with magnitude $H=2$--3~T\@.  Stars and asterisks represent the low temperature ($T\le 10$~K) values of $\chi_{ab}$ and $\chi_{c}$, respectively, extracted from $M(H)$ isotherms.  For the Sb-doped and Cr-doped crystals, the spurious noise and discontinuities in the magnitude and slope of the data for $H||c$ at about 140~K and 230~K, respectively, are due to a defect in the Quantum Design SQUID magnetometer magnetic moment evaluation software that occurs when the magnetic moment of the sample plus sample holder passes through zero (here the sample holder is diamagnetic and has been corrected for in the figure).  The anisotropic $\chi(T)$ data for a single crystal of ${\rm BaMn_2As_2}$ obtained by Singh et al.\ (Ref.~\onlinecite{YSingh1}) are included for comparison.}
	\label{fig:Figure_chi_T_All_Xtals}
\end{figure}

Shown in Fig.~\ref{fig:Figure_chi_T_All_Xtals} is the anisotropic temperature variation of the magnetic susceptibility $\chi \equiv M/H$ of the Ba(Mn$_{0.901}$Fe$_{0.099}$)$_2$As$_2$, ${\rm Ba(Mn_{0.956}Cr_{0.044})_2As_2}$ and ${\rm BaMn_2(As_{0.974}Sb_{0.026})_2}$ single crystals. The susceptibilities obtained as the slopes of the high-field linear part of the $M(H)$ isotherms for temperatures $\le 10$~K, in the field range of 3--5.5~T, are also plotted as distinct symbols in Fig.~\ref{fig:Figure_chi_T_All_Xtals}.  Similar to data for undoped BaMn$_2$As$_2$ (Ref.~\onlinecite{YSingh1}) that are also shown in the figure, the in-plane susceptibility $\chi_{ab} \sim 1.5\times 10^{-3}$~cm$^{3}$/mol is nearly independent of $T$ over the whole $T$ range of the measurement for all our crystals.  In contrast, the $c$~axis susceptibility $\chi_{c}$ generally decreases monotonically with decreasing $T$ below 300~K\@. These observations indicate that our substituted crystals have magnetic properties very similar to those of unsubstituted BaMn$_2$As$_2$ and therefore that the former substituted crystals are antiferromagnetically ordered with N\'eel temperatures $T_{\rm N}$ far above room temperature, where $T_{\rm N} = 625$~K for unsubstituted BaMn$_2$As$_2$.\cite{YSingh2}

\section{\label{SummConcl} Summary and Conclusions}

From our room-temperature powder XRD measurements of polycrystalline Ba(Mn$_x$Fe$_{1-x}$)$_2$As$_2$ samples slow-cooled from 1000\,$^\circ$C, we discovered a miscibility gap at 300~K with a composition range $0.12 \lesssim x \leq 1$.  However, the miscibility gap becomes narrower at higher temperatures.  For samples quenched into liquid nitrogen from 1000\,$^\circ$C, the miscibility gap narrows to the concentration region $0.2 \lesssim x \leq 0.8$.  The quenched samples of Ba(Mn$_x$Fe$_{1-x}$)$_2$As$_2$ with \emph{x} = 0.4, 0.5 and 0.6 contain a single 122-type phase but also contain significant amounts of impurities, mainly Fe$_{1-x}$Mn$_x$As and FeAs$_2$.  These latter results suggest that the equilibrium phase diagram between BaMn$_2$As$_2$ and BaFe$_2$As$_2$ is not pseudo-binary in nature near the middle of the composition range.

Electrical resistivity measurements suggest that our single-phase quenched polycrystalline Ba(Mn$_{0.8}$Fe$_{0.2}$)$_2$As$_2$ and Ba(Mn$_{0.2}$Fe$_{0.8}$)$_2$As$_2$ samples are not metallic, but we caution that reliable resistivity data on such materials can only be obtained using single crystals.  
Indeed, our heat capacity measurements on these two samples reveal significant linear terms at low temperatures, suggesting metallic ground states.  Our $M(H)$ measurements show that the quenched samples of Ba(Mn$_{0.8}$Fe$_{0.2}$)$_2$As$_2$ and Ba(Mn$_{0.2}$Fe$_{0.8}$)$_2$As$_2$ do not contain ferromagnetic impurities.  The magnitudes and temperature dependences of $\chi$ around room temperature for these compositions are qualitatively consistent with those reported for BaMn$_2$As$_2$ and BaFe$_2$As$_2$.  The resistivity data, but not the susceptibility data, for Ba(Mn$_{0.8}$Fe$_{0.2}$)$_2$As$_2$ suggest that some sort of phase transition may take place at about 260~K\@.  Furthermore, heat capacity data for both Ba(Mn$_{0.8}$Fe$_{0.2}$)$_2$As$_2$ and Ba(Mn$_{0.2}$Fe$_{0.8}$)$_2$As$_2$ show no evidence for any phase transitions from 2 to 280~K\@.  These $C_{\rm p}(T)$ data suggest that optical phonon excitations and/or magnetic excitations become important near and above room temperature.

Since the $\chi(T)$, $\rho(T)$ and $C_{\rm p}(T)$ data for Ba(Mn$_{0.2}$Fe$_{0.8}$)$_2$As$_2$ do not show any anomalies below 300~K, these data suggest that the substitution of 20\% of the Fe by Mn in BaFe$_2$As$_2$ suppresses the tetragonal-to-orthorhombic and SDW transitions that are observed in BaFe$_2$As$_2$.  However, the recent work of Kim et al.\ found that 17.6\% substitution of Fe by Mn suppresses the crystallographic transition, but not the SDW transition which still occurs at $T_{\rm N} \lesssim 200$~K.\cite{Kim2010b}  Furthermore, this study reported inflection points in the temperature dependences of the resistivities of their single crystal samples with $0\le x \le 0.176$ that occurred at the respective composition-dependent temperatures of the structural/magnetic transition.  Our resistivity measurement on polycrystalline Ba(Mn$_{0.2}$Fe$_{0.8}$)$_2$As$_2$ in Fig.~\ref{fig:Figure_7_Resistivity_AP1066}(b) did not show any evidence for such an inflection point.  It is not clear why our measurements on polycrystalline Ba(Mn$_{0.2}$Fe$_{0.8}$)$_2$As$_2$ do not show behaviors similar to those reported for single crystals at similar compositions.  

We attempted to grow single crystals of substituted BaMn$_2$As$_2$ out of Sn flux.  However, substitution of Mn by Co, Ni, Cu, Ru, Rh, Pd, Re and Pt did not occur in concentrations greater than 0.5\% except for 4.4\% and $\sim 10$\% substitutions in the case of Cr and Fe, respectively, and 2.6\% substitution of As by Sb.  The magnetic properties of Fe-, Cr- and Sb-substituted BaMn$_2$As$_2$ crystals are similar to those of pure BaMn$_2$As$_2$.

\acknowledgments

We are grateful to Yogesh Singh and Andreas Kreyssig for helpful discussions, Supriyo Das for assistance with laboratory activities, and Makariy Tanatar for advice on making electrical contacts to the samples for resistivity measurements.  Work at the Ames Laboratory was supported by the Department of Energy-Basic Energy Sciences under Contract No.~DE-AC02-07CH11358.

\end{document}